\documentstyle{article}
\begin{document}
\title{$N$ identical particles
under quantum confinement: A many-body dimensional perturbation
theory approach}
\author{B.\ A.\ McKinney\protect\footnote{Department of Molecular
Physiology and Biophysics, Vanderbilt University 
Medical School, Nashville, TN 37232.}, M.\ Dunn, D.\ K.\ Watson\\
University of Oklahoma \\ Department of Physics and Astronomy \\
Norman, OK 73019 \and J.\ G.\ Loeser \\
Oregon State University \\ Department of Chemistry \\ Corvallis,
OR 97331}
\date{\today}
\maketitle
\begin{abstract}
Systems that involve $N$ identical interacting particles under
quantum confinement appear throughout many areas of physics,
including chemical, condensed matter, and atomic physics.  In this
paper, we present the methods of dimensional perturbation theory,
a powerful set of tools that uses symmetry to yield simple results
for studying such many-body systems. We present a detailed
discussion of the dimensional continuation of the $N$-particle
Schr\"odinger equation, the spatial dimension $D\to\infty$
equilibrium ($D^0$)
structure, and the normal-mode ($D^{-1}$) structure.  We use the
FG matrix method to derive general, analytical expressions for the
many-body normal-mode vibrational frequencies, and we give
specific analytical results for three confined $N$-body quantum
systems: the $N$-electron atom, $N$-electron quantum dot, and
$N$-atom inhomogeneous Bose-Einstein condensate with a repulsive
hardcore potential.
\end{abstract}
\section{Introduction}
During the last two decades, techniques to confine and manipulate
atoms, ions, and electrons have led to the creation of new
$N$-body quantum systems of both fundamental and technological
interest. The condensation of atomic bose gases, the confinement
of atoms using optical lattices, and the creation of quantum dots
in semiconductors are all examples of novel many-body environments
that can be manipulated using external potentials. These systems
thus provide a unique opportunity to study many-body effects over
a range of interaction strengths as external potentials are tuned.
The quantum dot is a nanostructure in which electrons are confined
in all three dimensions by an external potential.  This
``artificial atom'' has shown great potential in biological and
medical applications and may form the basis for a new generation
of semiconductor lasers with applications in quantum computation
and quantum cryptography. The atomic vapor Bose-Einstein
condensate, arising from the collapse of a collection of
identical, harmonically trapped bosons into the lowest
single-particle state of the harmonic oscillator potential,
provides a source of coherent matter waves with possible
applications in atom interferometry, atom circuits, holography,
precision measurements, and quantum computation. The confinement
of atoms in single cells of optical lattices could provide the
controlled interaction required to create a quantum logic gate.

These new quantum confined $N$-body systems have resulted in renewed
interest and new demands in the many-body techniques of
quantum physics and chemistry, originally developed to study atoms
and molecules -- $N$-body systems confined by the mutual
attraction of the particles themselves. Mean-field treatments of
these $N$-body systems, such as the Hartree-Fock method in atomic
physics and the Gross-Pitaevskii approximation for Bose-Einstein
condensates, do not include many-body effects, and, therefore,
become inaccurate for systems under tight confinement or strong
interaction. These new systems, which can have a few hundred to
millions of particles, present serious challenges for existing
many-body methods, most of which were developed with small systems
in mind.

In this paper we offer the methods of many-body dimensional
perturbation theory for the study of large $N$-body systems under
quantum confinement. We apply dimensional perturbation theory
(DPT) to many-body quantum confined systems from chemical,
condensed matter, and atomic physics. The first application of
many-body DPT (Section \ref{sec:atom}), and the impetus for this
paper, originated in Ref.\cite{loeser}, in which Loeser introduced
low-order many-body dimensional perturbation methods to the
$N$-electron atom.  In this instrumental paper, Loeser obtains
low-order, analytical expressions for the ground-state energy of
neutral atoms. For $Z=1$ to $127$, the numerical results compare
well to Hartree-Fock energies with a correlation correction. In
Sections \ref{sec:SE}-\ref{sec:firstorder} of this paper, we
discuss the formalism and general theory behind Loeser's results
for the $N$-electron atom, in which the quantum confinement of the
$N$ electrons is supplied by the Coulomb attraction of a
nucleus. The second application is the quantum dot (Section
\ref{sec:qd}), an atom-like many-body system from condensed matter
physics, where the confinement of the $N$ electrons is supplied by
an external, isotropic trapping potential. The final system
(Section \ref{sec:hs}) is $N$ identical hard spheres in an
isotropic trap. Although our results are more general, this
hard-sphere model is appropriate for describing a Bose-Einstein
condensate of a trapped alkali-metal gas, a system that, like the
quantum dot, bridges the areas of atomic and condensed matter
physics.

In many-body dimensional perturbation theory, the coordinate
vectors of the $N$ particles are allowed to have $D$ Cartesian
components.  The singular limit $D \to \infty$ is taken as the
unperturbed system, and $1/D$ becomes the perturbation parameter.
The Schr\"odinger equation is written in terms of a similarity
transformed wave function that removes first-order derivatives
from the Laplacian and introduces a centrifugal-like potential
containing all of the explicit dimension dependence of the
Laplacian.  The explicit dimension-dependence of the
centrifugal-like potential is quadratic, and in order to
regularize the large-dimension limit of the Hamiltonian (the $D
\to \infty$ limit or the leading term in the $1/D$ expansion) we
choose a scaling of the length and energy that is also quadratic
in $D$. Then as $D\to\infty$ the second derivative parts of the
kinetic energy vanish and the particles become localized in the
bottom of an effective potential defined by a centrifugal-like
contribution from the kinetic energy and contributions from the
other potential energies (i.e., the confinement and interaction
potentials). In the case of attractive interparticle forces, the
repulsive centrifugal-like term stabilizes the large-$D$
configuration against collapse. The first-order quantum correction
corresponds to normal-mode vibrations about the large-$D$
effective potential minimum.  We find the normal-mode frequencies
using the FG matrix technique of molecular physics\cite{dcw}.
Higher orders can be calculated using a matrix method developed
specifically for dimensional perturbation
theory\cite{matrix_method}.

This paper is organized as follows.  In Section \ref{sec:SE} we
give the unscaled Schr\"odinger equation, and in Section
\ref{sec:infD} we discuss scaling methods and the zeroth-order ($D
\to \infty$) term of the $1/D$ expansion. In Section
\ref{sec:firstorder} we discuss the $1/D$ expansion and utilize
the FG matrix method, and we derive analytical expressions for the
normal-mode frequencies of vibration about the infinite-$D$
symmetric minimum, from which we derive the first-order energy
approximation. In Secs. \ref{sec:atom}-\ref{sec:hs}, we apply
these methods to the $N$-electron atom, $N$-electron quantum dot,
and $N$-atom Bose-Einstein condensate.

\section{$D$-dimensional $N$-body Schr\"odinger
equation}\label{sec:SE}  For an $N$-body system of particles
confined by a spherically symmetric potential and interacting via
a common two-body potential $g_{ij}$, the Schr\"odinger equation
in $D$-dimensional Cartesian coordinates is
\begin{equation}
\label{generalH} H \Psi = \left[ \sum\limits_{i=1}^{N} h_{i} +
\sum_{i=1}^{N-1}\sum\limits_{j=i+1}^{N} g_{ij} \right] \Psi = E
\Psi,
\end{equation}
where
\begin{equation} \label{eq:hi}
h_{i}=-\frac{\hbar^2}{2
m_{i}}\sum\limits_{\nu=1}^{D}\frac{\partial^2}{\partial
x_{i\nu}^2} +
V_{\mathtt{conf}}\left(\sqrt{\sum\nolimits_{\nu=1}^{D}x_{i\nu}^2}\right),
\end{equation}
\begin{equation}
\mbox{and} \;\;\;
g_{ij}=V_{\mathtt{int}}\left(\sqrt{\sum\nolimits_{\nu=1}^{D}\left(x_{i\nu}-x_{j\nu}
\right)^2}\right)
\end{equation}
are the single-particle Hamiltonian and the two-body interaction
potential, respectively. The operator $H$ is the $D$-dimensional
Hamiltonian, and $x_{i\nu}$ is the $\nu^{th}$ Cartesian component
of the $i^{th}$ particle. $V_{\mathtt{conf}}$ is the confining
potential. For the $N$-electron atom, $V_{\mathtt{conf}}$ is the
Coulomb attraction between the electrons and the nucleus, while
for the $N$-electron quantum dot and $N$-atom hard-sphere problem
we model the confinement as a harmonic trapping potential. The
two-body interaction potential $V_{\mathtt{int}}$ is Coulombic in
the first two systems and a hard sphere in the third.
\subsection{Transformation of the Laplacian}
Restricting our attention to spherically symmetric ($L=0$) states
of the many-body system, we transform from Cartesian
to internal coordinates. A convenient internal coordinate system
for confined systems is
\begin{equation}\label{eq:int_coords}
r_i=\sqrt{\sum_{\nu=1}^{D} x_{i\nu}^2} \;\;\; (1 \le i \le N)
\;\;\; \mbox{and} \;\;\;
\gamma_{ij}=cos(\theta_{ij})=\left(\sum_{\nu=1}^{D}
x_{i\nu}x_{j\nu}\right) / r_i r_j \;\;\; (1 \le i < j \le N),
\end{equation}
which are the $D$-dimensional scalar radii $r_i$ of the $N$
particles from the center of the confining potential and the
cosines $\gamma_{ij}$ of the $N(N-1)/2$ angles between the radial
vectors.  

Now for a function $\Psi$ dependent on only two functions $r(x)$
and $\gamma(x)$ one can write
\begin{equation}
\frac{d^2 \Psi(r(x),\gamma(x))}{ d x^2} = \frac{d^2 r}{d x^2}
\frac{d \Psi}{d r} + \frac{d^2 \gamma}{d x^2} \frac{d \Psi}{d
\gamma} + \left( \frac{d r}{d x} \right)^2 \frac{d^2 \Psi}{d r^2}
+ \left( \frac{d \gamma}{d x} \right)^2 \frac{d^2 \Psi}{d
\gamma^2} + 2 \frac{d \gamma}{d x} \frac{d r}{d x} \frac{d^2
\Psi}{d r d \gamma}.
\end{equation}
Generalizing this, when operating on the state
$\Psi(r_i(x_{i\nu}),\gamma_{i1}(x_{i\nu})\ldots
\gamma_{ik}(x_{i\nu}) \ldots  \gamma_{iN}(x_{i\nu}))$ (where
$k\not= i$ and $\nu=1,\ldots,D$)), $\nabla_i^2$ can be written in
terms of the internal coordinates of Eq. (\ref{eq:int_coords}) as
\begin{eqnarray}
\nabla_i^2\Psi\equiv\sum\limits_{\nu=1}^{D}\frac{\partial^2}{\partial
x_{i\nu}^2}\Psi  & = & \sum\limits_{\nu=1}^{D}\left(
\frac{\partial^2 r_i}{{\partial x_{i\nu}}^2}\right)
\frac{\partial}{\partial r_i}\Psi +
\sum\limits_{\nu=1}^{D}\sum\limits_{j\not= i} \left(
\frac{\partial^2 \gamma_{ij}}{{\partial x_{i\nu}}^2} \right)
\frac{\partial}{\partial \gamma_{ij}}\Psi +\nonumber\\
&&\sum\limits_{\nu=1}^{D} \left( \frac{\partial r_i}{\partial
x_{i\nu}} \right)^2 \frac{\partial^2}{{\partial r_i}^2}\Psi +
\sum\limits_{\nu=1}^{D}\sum\limits_{j\not= i}\sum\limits_{k\not=
i} \left( \frac{\partial \gamma_{ij}}{\partial x_{i\nu}} \right)
\left ( \frac{\partial \gamma_{ik}} {\partial x_{i\nu}} \right)
\frac{\partial^2}{\partial \gamma_{ij} \partial  \gamma_{ik}}\Psi + \\
&&2 \sum\limits_{\nu=1}^{D}\sum\limits_{j\not=i} \left(
\frac{\partial r_i}{\partial x_{i\nu}} \right) \left(
\frac{\partial\gamma_{ij}}{\partial x_{i\nu}} \right)
\frac{\partial^2}{\partial r_i \partial \gamma_{ij}}\Psi.\nonumber
\end{eqnarray}

The relevant derivatives of the internal coordinates are
\begin{eqnarray}
\frac{\partial r_i}{\partial x_{i\nu}} = \frac{x_{i\nu}}{r_i} &&
\frac{\partial \gamma_{ij}}{\partial x_{i\nu}} =
\frac{1}{r_i}\left( \frac{x_{j\nu}}{r_j} -
\frac{x_{i\nu}}{r_i}\gamma_{ij} \right) \\
\frac{\partial^2 r_i}{{\partial x_{i\nu}}^2} = \frac{1}{r_i}\left(
1 - \frac{x_{i\nu}^2}{r_i^2} \right) && \frac{\partial^2
\gamma_{ij}}{{\partial x_{i\nu}}^2} = \frac{1}{r_i^2} \left( 3
\frac{x_{i\nu}^2}{r_i^2}\gamma_{ij} - 2 \frac{x_{i\nu}}{r_i}
\frac{x_{j\nu}}{r_j}-\gamma_{ij}\right),
\end{eqnarray}
which lead to the effective $S$-wave Laplacian in internal
coordinates:
\begin{eqnarray}
\label{eq:laplacian} \sum\limits_{i}\nabla_i^2 \Psi & = &
\sum\limits_{i}\frac{D-1}{r_i}\frac{\partial}{\partial
r_i}\Psi-\sum\limits_{i}\frac{D-1}{r_i^2}\sum\limits_{j\not=
i}\gamma_{ij}\frac{\partial}
{\partial \gamma_{ij}}\Psi +\nonumber\\
&&\sum\limits_{i}\frac{\partial^2}{{\partial r_i}^2}\Psi
+\sum\limits_{i}\sum\limits_{j\not= i}\sum\limits_{k\not=
i}\frac{\gamma_{jk}-\gamma_{ij}\gamma_{ik}}
{r_i^2}\frac{\partial^2}{\partial \gamma_{ij} \partial
\gamma_{ik}}\Psi.
\end{eqnarray}
\subsection{Removal of first-order derivatives}
\label{sub:simtransf} Next we wish to find a transformation of the
Hamiltonian that removes the first-order derivatives from the
Laplacian (Eq. (\ref{eq:laplacian})) so that the kinetic energy
operator is reduced to a sum of terms of two kinds, namely, a
second-order derivative term and a repulsive centrifugal-like
term, which when attractive interparticle potentials are present,
stabilizes the system against collapse in the large-$D$ limit.
When this is done the zeroth and first orders of the dimensional
($1/D$) expansion of the Hamiltonian become exactly soluble for
any value of $N$. In the $D\to\infty$ limit, the second derivative
terms drop out, resulting in a static problem at zeroth order,
while first order corrections correspond to simple harmonic
normal-mode oscillations about the infinite-dimensional structure.

In Ref. \cite{avery}, Avery {\sl et al.} considered the problem of
performing a similarity transformation of the wave function $\Psi$
and operators $\widehat{O}$:
\begin{equation}\label{eq:simtransf}
\Phi = \chi^{-1} \Psi, \;\; \mbox{and} \;\;
\widetilde{O}=\chi^{-1} \widehat{O} \chi,
\end{equation}
where the transforming function, with adjustable parameters
$\alpha$ and $\beta$, is of the form:
\begin{equation}\label{eq:chi}
\chi = (r_1 r_2 \ldots r_N)^{-\alpha} \Gamma^{- \beta/2}.
\end{equation}
Here $\Gamma$ is the Gramian determinant, the determinant of the
matrix whose elements are $\gamma_{ij}$ (see Appendix
\ref{app:gram}). One of the cases considered by Avery {\sl et
al.}\cite{avery} for $\alpha$ and $\beta$ ($\alpha_1=(D-1)/2$ and
$\beta_1=(D-N-1)/2$) causes the weight function for matrix
elements, $W=J\chi^2$, to equal unity, where $J$ is the Jacobian
of the transformation to internal coordinates:
\begin{equation}
J = (r_1 r_2 \ldots r_N)^{D-1} \Gamma^{(D-N-1)/2}.
\end{equation}
As we discuss in another publication\cite{jacobian}, this scenario
has the advantage of making matrix elements easier to calculate
using DPT and makes the physical interpretation of the
large-dimension normal-mode structure more transparent. However,
the results in this paper are easier to derive if we use
$\alpha=\beta=(D-1)/2$, which removes the first-order derivatives
from the Laplacian while giving the same results as $\alpha_1$ and
$\beta_1$ through first order in $1/D$. Carrying out the
transformation of the Hamiltonian and wave function of Eq.
(\ref{generalH}) via Eq. (\ref{eq:simtransf}) with
$\alpha=\beta=(D-1)/2$ in Eq. (\ref{eq:chi}), the Schr\"odinger
equation becomes\cite{loeser}:
\begin{eqnarray}
&&{\displaystyle (T+V)\Phi = E \Phi} \label{eq:SE} \\
&&{\displaystyle T=\sum\limits_{i=1}^{N}\left(-\frac{\hbar^2}{2
m_i}\frac{\partial^2}{{\partial r_i}^2}-
\sum\limits_{j\not=i}\sum\limits_{k\not=i}\frac{\hbar^2(\gamma_{jk}-\gamma_{ij}
\gamma_{ik})}{2 m_i
r_i^2}\frac{\partial^2}{\partial\gamma_{ij}\partial\gamma_{ik}}
+\frac{\hbar^2(D- 1)(D-2N-1)}{8 m_i
r_i^2}\frac{\Gamma^{(i)}}{\Gamma} \right)} \label{eq:SE_T}\\
&&{\displaystyle V=\sum\limits_{i=1}^{N}V_{\mathtt{conf}}(r_i)+
\sum\limits_{i=1}^{N-1}\sum\limits_{j=i+1}^{N}
V_{\mathtt{int}}(r_{ij}).}
\end{eqnarray}
The Gramian matrix whose determinant is $\Gamma^{(i)}$ is the
$i^{th}$ principal minor formed by deleting from $\Gamma$ the row
and column corresponding to the $i^{th}$ particle. The quantity
$r_{ij}=\sqrt{r_{i}^2+r_{j}^2-2r_{i}r_{j}\gamma_{ij}}$ is the
interparticle separation.  The similarity transformed Hamiltonian
for the energy eigenstate $\Phi$ is $\chi^{-1} H \chi$, where
$H=(T+V)$.

We remark that for the hard-sphere system in Sec. \ref{sec:hs}, we
find it expedient to choose a dimensional continuation of
$V_{\mathtt{int}}$ that contains explicit dimension dependence,
which is not expressed in the equation above. However, the general
discussion to follow in Secs. \ref{sec:infD} -
\ref{sec:firstorder} holds for the hardsphere system with only
slight modification to the first-order energy approximation.  This
is discussed further in Sec. \ref{sec:hs}.

\section{Infinite-$D$ analysis: Leading order energy}\label{sec:infD}

To begin the perturbation analysis we regularize the
large-dimension limit of the Schr\"odinger equation by defining
dimensionally scaled variables:
\begin{equation}
\bar{r}_i = r_i/\kappa(D) \;\;\; , \;\;\; \bar{E} = \kappa(D) E
\;\;\; \mbox{and} \;\;\; \bar{H} = \kappa(D) H
\end{equation}
with dimension-dependent scale factor $\kappa(D)$.  From Eq.
(\ref{eq:SE_T}) one can see that the kinetic energy T scales in
the same way as $1/r^2$, so the scaled version of Eq.
(\ref{eq:SE}) becomes
\begin{equation}\label{eq:scale1}
\bar{H} \Phi = \left(\frac{1}{\kappa(D)}\bar{T}+\bar{V}
\right)\Phi = \bar{E} \Phi,
\end{equation}
where barred quantities simply indicate that the variables are now
in scaled units. Because of the quadratic $D$ dependence in the
centrifugal-like term in T of Eq. (\ref{eq:SE_T}), we conclude
that the scale factor $\kappa(D)$ must also be quadratic in $D$,
otherwise the $D\to\infty$ limit of the Hamiltonian would not be
finite. The factor of $\kappa(D)$ in the denominator of Eq.
(\ref{eq:scale1}) acts as an effective mass that increases with
$D$, causing the derivative terms to become suppressed while
leaving behind a centrifugal-like term in an effective potential,
\begin{equation}
\label{veff}
V_{\mathtt{eff}}=\sum\limits_{i=1}^{N}\left(\frac{\hbar^2}{8 m_i
\bar{r}_i^2}\frac{\Gamma^{(i)}}{\Gamma}+V_{\mathtt{conf}}(\bar{r}_i)\right)+\sum\limits_{i=1}^{N-1}\sum\limits_{j=i+1}^{N}
V_{\mathtt{int}}(\bar{r}_{ij}),
\end{equation}
in which the particles become frozen at large $D$. In the
$D\to\infty$ limit, the excited states have collapsed onto the
ground state, which is found at the minimum of $V_{\mathtt{eff}}$.

We assume a totally symmetric minimum characterized by the
equality of all radii and angle cosines of the particles when
$D\to\infty$, i.e.,
\begin{equation}
\bar{r}_{i}=\bar{r}_{\infty} \;\; (1 \le i \le N), \;\;\;\;
\gamma_{ij}=\gamma_{\infty} \;\; (1 \le i < j \le N).
\end{equation}
Since each particle radius and angle cosine is equivalent, we can
take derivatives with respect to an arbitrary $\bar{r}_{i}$ and
$\gamma_{ij}$ in the minimization procedure. Then evaluating all
$\bar{r}_{i}$ and $\gamma_{ij}$ at the infinite-$D$ radius and
angle cosine, $\bar{r}_{\infty}$ and $\gamma_{\infty}$,
respectively, we find that $\bar{r}_{\infty}$ and
$\gamma_{\infty}$ satisfy
\begin{eqnarray}
\label{minimum1} \frac{\partial V_{\mathtt{eff}}}{\partial
\bar{r}_{i}}\Biggr|_{\infty}&=&0
\\ \label{minimum2} \frac{\partial V_{\mathtt{eff}}}{\partial
\gamma_{ij}}\Biggr|_{\infty}&=&0,
\end{eqnarray}
where the $\infty$ subscript means to evaluate all $\bar{r}_{i}$
at $\bar{r}_{\infty}$ and all $\gamma_{ij}$ at $\gamma_{\infty}$.
In scaled units the zeroth-order ($D\to\infty$) approximation for
the energy becomes
\begin{equation}
\label{zeroth}
\bar{E}_{\infty}=V_{\mathtt{eff}}\Big|_{\infty}=V_{\mathtt{eff}}(\bar{r}_{\infty},\gamma_{\infty}).
\end{equation}
In this leading order approximation, the centrifugal-like term
that appears in $V_{\mathtt{eff}}$, even for the ground state, is
a zero-point energy contribution required by the minimum
uncertainty principle\cite{chat}.
\section{Normal-mode analysis: $1/D$ first-order quantum energy correction}\label{sec:firstorder}

At zeroth-order, the particles can be viewed as frozen in a
completely symmetric, high-$D$ configuration or simplex, which is
somewhat analogous to the Lewis structure in atomic physics
terminology. Likewise, the first-order $1/D$ correction can be
viewed as small oscillations of this structure, analogous to
Langmuir oscillations. Solving Eqs. (\ref{minimum1}) and
(\ref{minimum2}) for $\bar{r}_{\infty}$ and $\gamma_{\infty}$
gives the infinite-$D$ structure and zeroth-order energy and
provides the starting point for the $1/D$ expansion. To obtain the
$1/D$ quantum correction to the energy for large but finite values
of $D$, we expand about the minimum of the $D\to\infty$ effective
potential. We first define a position vector, consisting of all
$N(N+1)/2$ internal coordinates, whose transpose is:
\begin{equation}\label{eq:ytranspose}
{\bar{y}}^{T} =
(\bar{r}_1,\bar{r}_2,\ldots,\bar{r}_N,\gamma_{12},\gamma_{13},\ldots,\gamma_{N-1,N}).
\end{equation}
We then make the following substitutions for all radii and angle
cosines:
\begin{eqnarray}
\label{eq:taylor1}
&&\bar{r}_{i} = \bar{r}_{\infty}+\delta^{1/2}\bar{r}'_{i}\\
&&\gamma_{ij} = \gamma_{\infty}+\delta^{1/2}\gamma'_{ij}
\label{eq:taylor2},
\end{eqnarray}
where $\delta=1/D$ is the expansion parameter, and we define a
displacement vector consisting of the internal displacement
coordinates [primed in Eqs. (\ref{eq:taylor1}) and
(\ref{eq:taylor2})]:
\begin{equation}\label{eq:ytransposeP}
\bar{y}'^{T} =
(\bar{r}'_{1},\bar{r}'_{2},\ldots,\bar{r}'_{N},\gamma'_{12},\gamma'_{13},\ldots,\gamma'_{N-1,N}).
\end{equation}
We may then obtain a power series in $\delta^{1/2}$ of the
effective potential about the $D\to\infty$ symmetric minimum:
\begin{equation}
\label{Taylor} V_{\mathtt{eff}}=V_{\mathtt{eff}}\Big|_{\infty} +
\delta^{1/2} \sum\limits_{\sigma=1}^{P}\left[\frac{\partial
V_{\mathtt{eff}}}{\partial \bar{y}'_{\sigma}}\right]_{\infty}
\bar{y}'_{\sigma} + \frac{1}{2} \delta \sum\limits_{\sigma=1}^{P}
\sum\limits_{\nu=1}^{P} \bar{y}'_{\sigma} \left[\frac{\partial^2
V_{\mathtt{eff}}}{\partial \bar{y}'_{\sigma} \partial
\bar{y}'_{\nu}}\right]_{\infty} \bar{y}'_{\nu} +
O\left(\delta^{3/2}\right),
\end{equation}
where
\begin{equation}
P \equiv N(N+1)/2
\end{equation}
is the number of internal coordinates. The first term in the power
series (Eq. (\ref{Taylor})) is simply the zeroth-order energy (Eq.
(\ref{zeroth})). The second term is zero since we are expanding
about the minimum of the effective potential; the system is said
to be in equilibrium since the forces acting on the system vanish
[Eqs. (\ref{minimum1}) and (\ref{minimum2})]. The third term
defines the elements of the Hessian matrix\cite{strang} ${\bf F}$
of Eq. (\ref{Gham}) below. The derivative terms in the kinetic
energy are taken into account by a similar series expansion,
beginning with a first-order term that is bilinear in
${\partial/\partial \bar{y}'}$, i.e.,
\begin{equation}\label{eq:T}
{\mathcal T}=-\frac{1}{2} \delta \sum\limits_{\sigma=1}^{P}
\sum\limits_{\nu=1}^{P} {G}_{\sigma\nu}
\partial_{\bar{y}'_{\sigma}}
\partial_{\bar{y}'_{\nu}} + O\left(\delta^{3/2}\right),
\end{equation}
where ${\mathcal T}$ is the derivative portion of the kinetic
energy $T$ [see Eq. (\ref{eq:SE_T})].  Thus, obtaining the
first-order energy correction is reduced to a harmonic problem,
which is solved by obtaining the normal modes of the system.

We use the Wilson FG matrix method\cite{dcw} to obtain the
normal-mode vibrations and, thereby, the first-order energy
correction. It follows from Eqs. (\ref{Taylor}) and (\ref{eq:T})
that ${\bf G}$ and ${\bf F}$, both constant matrices, are defined
in the first-order $\delta=1/D$ Hamiltonian as follows:
\begin{equation}\label{Gham}
\widehat{H}_1=-\frac{1}{2} {\partial_{\bar{y}'}}^{T} {\mathbf G}
{\partial_{\bar{y}'}} + \frac{1}{2}\bar{y}'^{T} {\mathbf F}
{\bar{y}'}.
\end{equation}
After the Schr\"odinger equation (\ref{eq:SE}) has been
dimensionally scaled, the second-order derivative terms are of
order $\delta$, and by comparing these terms with the first part
of $\widehat{H}_1$, the elements of the kinetic-energy matrix
${\bf G}$ are easily determined.  The elements of the Hessian
matrix\cite{strang}, ${F}_{\sigma\nu}=[\partial^2
V_{\mathtt{eff}}/\partial \bar{y}'_{\sigma}
\bar{y}'_{\nu}]_{\infty}$, on the other hand, require a bit more
effort to obtain, as we will see in detail in later sections.

We include a derivation of the FG matrix method in Appendix
\ref{app:wilson}, but we state here the main result of the method,
which consists of finding the roots of the following
characteristic polynomial in $\lambda$:
\begin{equation}\label{eq:character}
\det(\lambda{\bf I}-{\bf G}{\bf F})=0.
\end{equation}
Depending on the number of particles, the number of roots
$\lambda$ -- there are $P \equiv N(N+1)/2$ roots -- is potentially
quite large. However, as we will see shortly, there is a high
degree of degeneracy due to the total symmetry of the infinite-$D$
Lewis structure.  In fact, there are only five distinct roots,
$\lambda_{\mu}$, where $\mu$ runs over ${\bf 0}^-$, ${\bf
0}^+$, ${\bf 1}^-$, ${\bf 1}^+$, and ${\bf 2}$.  And as we
conclude in Eq. (\ref{eq:E1}), the energy through first-order can
be written in terms of the distinct normal-mode vibrational
frequencies, which are related to the roots $\lambda_{\mu}$ of
${\bf GF}$ by
\begin{equation}\label{eq:omega_p}
\lambda_{\mu}=\bar{\omega}_{\mu}^2,
\end{equation}
as can be seen from Eq. (\ref{eq:appH1}) in Appendix
\ref{app:wilson}. In the next subsections, we show explicitly how
to find analytical expressions for the roots of ${\bf GF}$.

\subsection{Indical structure of F, G, and GF matrices}

The ${\bf F}$, ${\bf G}$, and ${\bf GF}$ matrices, which we
generically denote by ${\bf Q}$, are $P \times P$ matrices with
the same indical structure as $\bar{y} \bar{y}^T$:
\begin{equation}\label{eq:yTy}
\bar{y}\bar{y}^T=\left(
\begin{array}{ccccccccc}
\bar{r}_1\bar{r}_1     &\bar{r}_1\bar{r}_2  &\cdots                &\bar{r}_1\bar{r}_N     &\vline&\bar{r}_1\gamma_{12}&\bar{r}_1\gamma_{13}&\cdots&\bar{r}_1\gamma_{N-1 N}\\
\bar{r}_2\bar{r}_1     &\bar{r}_2\bar{r}_2  &\cdots                &                       &\vline&\bar{r}_2\gamma_{12}&\bar{r}_2\gamma_{13}&\cdots&\bar{r}_2\gamma_{N-1 N} \\
\vdots                 &\vdots              &\ddots                &    \vdots             &\vline&\vdots              &\vdots              & \ddots          &\vdots \\
\bar{r}_N\bar{r}_1     &\cdots              &                      &\bar{r}_N\bar{r}_N     &\vline&\bar{r}_N\gamma_{12}& &\cdots&\bar{r}_N\gamma_{N-1 N} \\
\hline
\gamma_{12}\bar{r}_1   &\gamma_{12}\bar{r}_2& \cdots               &\gamma_{12}\bar{r}_N   &\vline&\gamma_{12}\gamma_{12}&\gamma_{12}\gamma_{13}&\cdots&\gamma_{12}\gamma_{N-1 N}\\
\gamma_{13}\bar{r}_1   &\gamma_{13}\bar{r}_2& \cdots               &\gamma_{12}\bar{r}_1   &\vline&\gamma_{13}\gamma_{12}&\gamma_{13}\gamma_{13}&\cdots&\gamma_{13}\gamma_{N-1 N}\\
\vdots                 &\vdots              & \ddots               &\vdots                 &\vline&\vdots                &\vdots                &\ddots&\vdots \\
\gamma_{N-1 N}\bar{r}_1& & \cdots& \gamma_{N-1 N}\bar{r}_N &\vline
&\gamma_{N-1 N}\gamma_{12}&\gamma_{N-1 N}\gamma_{13}
&\cdots&\gamma_{N-1 N}\gamma_{N-1 N}
\end{array}
\right),
\end{equation}
where $\bar{y}$ is defined by Eq. (\ref{eq:ytranspose}). The
indical structure of this matrix suggests a convenient shorthand
for referencing the elements of the ${\bf Q}$ matrices. The upper
left block of Eq. (\ref{eq:yTy}) is an $(N \times N)$ matrix with
elements associated with $(\bar{r}_i,\bar{r}_j)$; hence we use the
subscript $(i,j)$ to refer to these elements. The upper right
block is an $(N \times N(N-1)/2)$ matrix with elements associated
with $(\bar{r}_i, \gamma_{jk})$; hence, we use the subscript
$(i,jk)$ to refer to these elements. The lower left block is an
$(N(N-1)/2 \times N)$ matrix with elements associated with
$(\gamma_{ij},\bar{r}_k)$; hence, we use the subscript $(ij,k)$ to
refer to these elements. Finally, the lower right block is an
$(N(N-1)/2 \times N(N-1)/2)$ matrix with elements associated with
$(\gamma_{ij},\gamma_{kl})$; hence, we use the subscript $(ij,kl)$
to refer to these elements.

\subsection{Symmetry of the Q matrices}
As the number of particles $N$ increases, diagonalizing the $P
\times P$ ${\bf GF}$ matrix (where $P \equiv N(N+1)/2$) becomes,
prima facie, a daunting task. However, one of the advantages of
dimensional perturbation theory is the simplifications that occur
in the large-dimension limit. In particular, since we are dealing
with identical particles in a totally symmetric configuration (the
Lewis structure) in which all the particles are equivalent, the
${\bf Q}$ matrices display a high degree of symmetry with many
identical elements.  Specifically,
\begin{equation}\label{eq:GFsyma}
\begin{array}{llllll}
Q_{i,i}&=&Q_{i',i'} &\equiv& Q_a   & \\
Q_{i,j}&=&Q_{i',j'} &\equiv& Q_b   & (i \neq j) \;\; \mbox{and} \;\; (i' \neq j') \\
Q_{ij,i}&=&Q_{i'j',i'} &\equiv& Q_c & (i \neq j) \;\; \mbox{and} \;\; (i' \neq j')\\
Q_{jk,i}&=&Q_{j'k',i'} &\equiv& Q_d & (i \neq j \neq k) \;\; \mbox{and} \;\; (i' \neq j' \neq k')\\
Q_{i,ij}&=&Q_{i',i'j'} &\equiv& Q_e & (i \neq j) \;\; \mbox{and} \;\; (i' \neq j') \\
Q_{i,jk}&=&Q_{i',j'k'} &\equiv& Q_f & (i \neq j \neq k) \;\; \mbox{and} \;\; (i' \neq j' \neq k')\\
Q_{ij,ij}&=& Q_{i'j',i'j'} &\equiv& Q_g & (i \neq j) \;\; \mbox{and} \;\; (i' \neq j')\\
Q_{ij,jk}&=& Q_{i'j',j'k'} &\equiv& Q_h & (i \neq j \neq k) \;\; \mbox{and} \;\; (i' \neq j' \neq k')\\
Q_{ij,kl}&=&Q_{i'j',k'l'} &\equiv& Q_{\iota} & (i \neq j \neq k
\neq l) \;\; \mbox{and} \;\; (i' \neq j' \neq k' \neq l').
\end{array}
\end{equation}
Note the indices in the relationships above run over all particles
$(1,2,\ldots,N)$ with the exceptions noted in the far right
column.  For example, $Q_{i,j}=Q_{i',j'}\equiv Q_b$, where $(i
\neq j)$ and $(i' \neq j')$, means that all off-diagonal elements
of the upper left block (the pure radial block) of ${\bf Q}$ are
equal to the same constant $Q_b$. Similarly,
$Q_{ij,kl}=Q_{i'j',k'l'}\equiv Q_{\iota}$, where $(i \neq j \neq k
\neq l)$ and $(i' \neq j' \neq k' \neq l')$, means that any
elements of ${\bf Q}$ in the lower right block (the pure angular
block) that do not have a repeated index are all equal to the same
constant $Q_{\iota}$.  We should remark here that ${\bf G}$ and
${\bf F}$ are also symmetric matrices (${\bf G}^T={\bf G}$ and
${\bf F}^T={\bf F}$); however, while ${\bf GF}$ does display the
high degree of symmetry of Eq. (\ref{eq:GFsyma}), it is not a
symmetric matrix.

\subsection{Q matrices in terms of simple submatrices}

The symmetry of the ${\bf Q}$ matrices (${\bf F}$, ${\bf G}$, and
${\bf GF}$) described in Eq. (\ref{eq:GFsyma}) allows us to write
these matrices in terms of six simple submatrices. We first define
the number of $\gamma_{ij}$ coordinates to be
\begin{equation}\label{eq:M}
M \equiv N(N-1)/2,
\end{equation}
and let ${\bf I}_N$ be an $N \times N$ identity matrix, ${\bf
I}_M$ an $M \times M$ identity matrix, ${\bf J}_N$ an $N \times N$
matrix of ones and ${\bf J}_M$ an $M \times M$ matrix of ones.
Further, we let ${\bf R}$ be an $N \times M$ matrix\footnote{In
graph theoretic terminology ${\bf R}$ is called a vertex-edge
matrix (see Appendix \ref{app:graph}).} such that
${R}_{i,jk}=\delta_{ij}+\delta_{ik}$, ${\bf J}_{NM}$ be an $N
\times M$ matrix of ones, and ${\bf J}^T_{NM}={\bf J}_{MN}$.

We then write the ${\bf Q}$ matrices as
\begin{equation}\label{eq:Q}
{\bf Q}=\left(\begin{array}{cc} {\bf Q}_1 & {\bf Q}_2 \\ {\bf Q}_3
& {\bf Q}_4
\end{array}\right),
\end{equation}
where the block ${\bf Q}_1$ has dimension $(N \times N)$, block
${\bf Q}_2$ has dimension $(N \times M)$, block ${\bf Q}_3$ has
dimension $(M \times N)$, and block ${\bf Q}_4$ has dimension $(M
\times M)$.  Now, as we show in Appendix \ref{app:graph}, Eq.
(\ref{eq:GFsyma}) allows us to write the following:
\begin{eqnarray}
{\bf Q}_1&=&(Q_a-Q_b) {\bf I}_N + Q_b {\bf J}_N \\
{\bf Q}_2&=&(Q_e-Q_f) {\bf R} + Q_f {\bf J}_{NM}\\
{\bf Q}_3&=&(Q_c-Q_d) {\bf R}^T + Q_d {\bf J}_{NM}^T\\
{\bf Q}_4&=&(Q_g-2Q_h+Q_{\iota}) {\bf I}_M + (Q_h-Q_{\iota}) {\bf
R}^T {\bf R} + Q_{\iota} {\bf J}_M.
\end{eqnarray}
In particular, letting ${\bf Q}={\bf GF}$, the matrix that must be
diagonalized, Eq. (\ref{eq:Q}) becomes
\begin{equation}\label{GFsub}
{\bf GF}=\left(
\begin{array}{cc}
(a-b) {\bf I}_N + b {\bf J}_N & (e-f) {\bf R} + f {\bf J}_{NM} \\
(c-d) {\bf R}^T + d {\bf J}_{MN} & (g-2h+\iota) {\bf I}_M +
(h-\iota) {\bf R}^T {\bf R} + \iota {\bf J}_M
\end{array}\right),
\end{equation}
where we have used the following abbreviations:
\begin{eqnarray}\label{GFsym}
a&\equiv&({GF})_{a}={G}_{a}{F}_{a}   \nonumber \\
b&\equiv&({GF})_{b}={G}_{a}{F}_{b} \nonumber\\
c&\equiv&({GF})_{c}={G}_{g}{F}_{e} +
(N-2){G}_{h}{F}_{e} + (N-2){G}_{g}{F}_{f} \nonumber\\
d&\equiv&({GF})_{d}={G}_{g}{F}_{f} +
2{G}_{h}{F}_{e} + 2(N-3){G}_{h}{F}_{f}  \nonumber\\
e&\equiv&({GF})_{e}={G}_{a}{F}_{e}  \\
f&\equiv&({GF})_{f}={G}_{a}{F}_{f}  \nonumber \\
g&\equiv&({GF})_{g}={G}_{g}{F}_{g}+2(N-2){G}_{h}{F}_{f} \nonumber\\
h&\equiv&({GF})_{h}={G}_{g}{F}_{h}+{G}_{h}{F}_{e}+(N-2){G}_{h}{F}_{h}
+(N-3){G}_{h}{F}_{\iota} \nonumber\\
\iota&\equiv&({GF})_{\iota}={G}_{g}{F}_{\iota}+4{G}_{h}{F}_{h}+2(N-4){G}_{h}{F}_{\iota}.
\nonumber
\end{eqnarray}
The right-hand sides of Eq. (\ref{GFsym}), the ${\bf GF}$ matrices
expressed in terms of the ${\bf F}$ and ${\bf G}$ matrix elements,
are derived in Appendix \ref{app:graph}.

\subsection{Normal mode frequencies and first-order energy}\label{section:det_E}
Having obtained the {\bf GF} matrix of Eqs. (\ref{GFsub}) and
(\ref{GFsym}), we may then, according to Eq. (\ref{eq:character}),
find the eigenvalues $\lambda$ of ${\bf GF}$ by solving
\begin{equation}
\det({\bf E})=0,
\end{equation}
where we have defined ${\bf E}$ as
\begin{equation}\label{eq:ematrix} {\bf E} \equiv  \left(
\begin{array}{cc}
\lambda {\bf I}_N & {\bf 0}\\ {\bf 0} & \lambda {\bf I}_M
\end{array}\right) - {\bf GF}.
\end{equation}
To find an analytical expression for $\det({\bf E})$ we multiply
${\bf E}$ by three matrices: ${\bf X}$, ${\bf Y}$ and ${\bf Z}$,
such that
\begin{equation}
\label{xyez} {\bf XYEZ} = \left(\begin{array}{cc} \frac{t}{v} {\bf
I}_N + \frac{u}{v} {\bf J}_N & {\bf 0} \\ \cdots & v {\bf I}_M
\end{array}\right),
\end{equation}
where $\det({\bf X})=\det({\bf Y})=\det({\bf Z})=1$ so that
$\det({\bf E})=\det({\bf XYEZ})$. In Appendix \ref{app:xyez} we
construct the matrices ${\bf X}$, ${\bf Y}$ and ${\bf Z}$ needed
to transform ${\bf E}$ of Eq. (\ref{eq:ematrix}) to ${\bf XYEZ}$
of Eq. (\ref{xyez}). We find
\begin{eqnarray}
v &\equiv& \lambda-g+2h-\iota \label{eq:v}\\
t&\equiv&(\lambda-a+b)v + (N-2)(\iota-h)(\lambda-a+b) +
(N-2)(d-c)(e-f) \label{eq:t}
\\ u &\equiv&
-kv-(h-\iota)(\lambda-a+b)+(d-c)(e-f) \nonumber \\
&\; \;& +\frac{N-1}{2}
\left[f(2(d-c)-Nd)-\iota(\lambda-a-(N-1)b)+4b(h-\iota)-2d(e-f)\right]
\label{eq:u}.
\end{eqnarray}

The determinant of ${\bf XYEZ}$, which does not depend on the
submatrix in the lower left block, is found from Appendix
\ref{app:graph} [see Eq. (\ref{PKn})] to be
\begin{equation}\label{eq:detalg}
\det({\bf E})=\det({\bf XYEZ})= t^{N-1} (t + N u) v^{M-N}.
\end{equation}
There are five distinct roots, which naturally arise from the
terms in Eq. (\ref{eq:detalg}) with $t$, $u$, and $v$ given by
Eqs. (\ref{eq:v}), (\ref{eq:t}) and (\ref{eq:u}). From Eq.
(\ref{eq:detalg}), $\det({\bf E})$ factors into three terms,
resulting in three equations for the five distinct roots:
\begin{eqnarray}
t^{N-1}=0 \label{troot}\\
(t+Nu)=0\\
v^{M-N}=v^{N(N-3)/2}=0\label{vroot},
\end{eqnarray}
From Eqs. (\ref{eq:v}) and (\ref{eq:t}), one can see that $t$ is
quadratic in $\lambda$. Hence, from Eq. (\ref{troot}) there are
two roots, which we denote by ${\bf 1}^-$ and ${\bf 1}^+$, with
multiplicity $d_{{\bf 1}^-} = d_{{\bf 1}^+} = N-1$ given by
\begin{equation}\label{teq0}
t=0.
\end{equation}
Similarly, the $(t+Nu)$ term in Eq. (\ref{eq:detalg}) is quadratic
in $\lambda$ and, hence, there are two roots, which we
denote by ${\bf 0}^-$ and ${\bf 0}^+$, with multiplicity
$d_{{\bf 0}^-} = d_{{\bf 0}^+} = 1$. Finally, $v$ is linear
in $\lambda$; hence, there is one root, which we denote by ${\bf 2}$,
with multiplicity $d_{{\bf 2}} = N(N-3)/2$ from Eq. (\ref{vroot}) given
 by
\begin{equation}\label{veq0}
v=0.
\end{equation}
 From Eqs. (\ref{veq0}) and (\ref{eq:v}), the root of multiplicity
$N(N-3)/2$ is
\begin{equation}\label{eq:charpol2}
\lambda_2=g-2h+\iota.
\end{equation}
From Eqs. (\ref{teq0}) and (\ref{eq:t}), the two roots with
multiplicity $(N-1)$ are
\begin{equation}\label{eq:charpol1}
\lambda_{1^{\pm}}=\eta_1 \pm \sqrt{{\eta_1}^2-\Delta_1},
\end{equation}
where
\begin{equation}\label{lam1defs}
\begin{array}{l}
\eta_1 = \frac{1}{2}\left[a-b+g+(N-4)h-(N-3)\iota\right] \\
\Delta_1 = (N-2)(c-d)(e-f)+(a-b)\left[g+(N-4)h-(N-3)\iota\right].
\end{array}
\end{equation}
From the $t+Nu$ term in Eq. (\ref{eq:detalg}) along with Eqs.
(\ref{eq:t}) and (\ref{eq:u}), the two roots with unit
multiplicity are
\begin{equation}\label{eq:charpol0}
\lambda_{0^{\pm}}=\eta_0 \pm \sqrt{{\eta_0}^2-\Delta_0},
\end{equation}
where
\begin{equation}\label{lam0defs}
\begin{array}{l}
\eta_0 = \frac{1}{2}\left[a-(N-1)b+g+2(N-2)h+\frac{(N-2)(N-3)}{2}\iota\right] \\
\Delta_0 =
(a-(N-1)b)\left[g+2(N-2)h-\frac{(N-2)(N-3)}{2}\iota\right]-\frac{N-2}{2}(2c+(N-2)d)(2e+(N-2)f).
\end{array}
\end{equation}

The five distinct roots $\lambda_{\mu}$, where $\mu$ runs
over ${\bf 0}^-$, ${\bf 0}^+$, ${\bf 1}^-$, ${\bf 1}^+$, and ${\bf
2}$, are equal to the square of the normal-mode vibrational
frequencies (i.e., Eq. (\ref{eq:omega_p})), and the energy through
first-order in $\delta=1/D$ is then
\begin{eqnarray}
\overline{E} &=& \overline{E}_{\infty} + \delta \overline{E}_o + O(\delta^2) \nonumber \\
&=&V_{\mathtt{eff}}(\bar{r}_{\infty},\gamma_{\infty}) +
\hspace{0.50em} \delta
\hspace{-2.00em}
\sum_{\renewcommand{\arraystretch}{0}
\begin{array}[t]{r@{}l@{}c@{}l@{}l} \scriptstyle \mu = \{
  & \scriptstyle \mathbf{0}^\pm,\hspace{0.5ex}
  & \scriptstyle \mathbf{1}^\pm & , & \\
  & & \scriptstyle \mathbf{2} & & \scriptstyle  \}
            \end{array}
            \renewcommand{\arraystretch}{1} }
\hspace{-0.50em}
\sum_{\mathsf{n}_{\mu}=0}^\infty ({\mathsf{n}}_{\mu}+\frac{1}{2})
d_{\mu,\mathsf{n}_{\mu}} \bar{\omega}_{\mu} +
O(\delta^2)\,, \label{eq:E1}
\end{eqnarray}
where the $\mathsf{n}_{\mu}$ are the vibrational quantum numbers
of the normal modes of the same frequency $\bar{\omega}_{\mu}$
(as such, $\mathsf{n}_{\mu}$ counts the number of nodes in a
given normal mode).
The quantity $d_{\mu,\mathsf{n}_{\mu}}$ is the occupancy of
the manifold of normal modes with vibrational quantum number
$\mathsf{n}_{\mu}$ and normal mode frequency
$\bar{\omega}_{\mu}$, i.e.\ it is the number of normal modes with the same
frequency $\bar{\omega}_{\mu}$ and the same number of quanta
$\mathsf{n}_{\mu}$.  The total occupancy of the
normal modes with frequency $\bar{\omega}_{\mu}$ is equal to
the multiplicity of the root $\lambda_{\mu}$ (see the
discussion after Eq. (\ref{vroot})), i.e.\
\begin{equation}
d_{\mu} = \sum_{\mathsf{n}_{\mu}=0}^\infty
            d_{\mu,\mathsf{n}_{\mu}} \,,
\end{equation}
where $d_{\mu}$ is the multiplicity of the $\mu^{th}$ root.
Note that although the equation in Ref.~\cite{loeser} for the energy
through Langmuir order is the same as Eq. (\ref{eq:E1}),
it is expressed a little differently (See Ref. \cite{different}).

\section{$N$-electron atom}
\label{sec:atom} The $N$-electron atom was previously discussed
using low-order many-body dimensional perturbation theory by
Loeser\cite{loeser,loeser2}. In this section, we give details
necessary to derive the results in Ref. \cite{loeser}, to which we
direct the interested reader for specific numerical results.  The
confining potential is provided by the Coulomb attraction of the
electrons to the nucleus:
\begin{equation}
V_{\mathtt{conf}}(r_i)=-\frac{Z}{r_i},
\end{equation}
and the interaction potential is the Coulomb repulsion of the
electrons:
\begin{equation}
V_{\mathtt{int}}(r_{ij})=\frac{1}{\sqrt{r_i^2 + r_j^2 - 2 r_i r_j
\gamma_{ij}}}.
\end{equation}

Substitution of the following charge/dimensionally scaled
variables into the similarity transformed Schr\"odinger equation
(\ref{eq:SE}) in atomic units ($\hbar=m=1$):
\begin{equation}
\bar{r}_i = \frac{Z}{\Omega}r_i, \;\;\;\;\; \bar{E} =
\frac{\Omega}{Z^2}E, \;\; \mbox{and} \;\; \Omega=(D-1)(D-2N-1)/4,
\end{equation}
places all of the dimension dependence in the second-derivative
parts of the kinetic energy and gives the following scaled
equation:
\begin{equation}
\left({\mathcal
T}^{(\mathtt{A})}+U^{(\mathtt{A})}+V^{(\mathtt{A})}\right)\Phi =
\bar{E} \Phi,
\end{equation}
where
\begin{eqnarray}\label{eq:resc_hamilt}
&& {\mathcal
T}^{(\mathtt{A})}=-\frac{1}{2\Omega}\sum\limits_{i=1}^N \left[
\frac{\partial^2}{{\partial\bar{r}_i}^2}+
\sum\limits_{j\not=i}\sum\limits_{k\not=i}\frac{\gamma_{jk}-\gamma_{ij}
\gamma_{ik}}{{\bar{r}_i}^2}\frac{\partial^2}{\partial\gamma_{ij}\partial\gamma_{ik}}
\right] \nonumber\\
&&U^{(\mathtt{A})}=\frac{1}{2}\sum\limits_{i=1}^{N}\frac{1}{{\bar{r}_i}^2}\frac{\Gamma^{(i)}}{
\Gamma} \\
&&V^{(\mathtt{A})}=-\sum\limits_{i=1}^{N}\frac{1}{\bar{r}_i}+\frac{1}{Z}
\sum\limits_{i=1}^{N-1}\sum\limits_{j=i+1}^{N}\frac{1}{\sqrt{{\bar{r}_i}^2+
{\bar{r}_j}^2-2\bar{r}_i\bar{r}_j\gamma_{ij}}}.\nonumber
\end{eqnarray}
The superscript A refers to the $N$-electron atom system (to
distinguish these quantities from those in the systems to follow).
From this equation, it can be seen that if $D$, and hence
$\Omega$, become infinitely large, the differential part
${\mathcal T}^{(\mathtt{A})}$ of the kinetic energy
($U^{(\mathtt{A})}$ being the centrifugal-like part of the kinetic
energy) will drop out of the similarity transformed Hamiltonian.
In effect it is as though the particles become infinitely heavy,
and, because of this, any function from the basis set of all delta
functions in configuration space is an eigenfunction of the
Hamiltonian. The energy of such an eigenfunction is just the value
of the effective potential at that specific point in configuration
space which that delta function selects, and, as mentioned in
Section \ref{sec:infD}, this point is the minimum of the
infinite-$D$ effective potential:
\begin{equation}
\bar{E}^{(\mathtt{A})}_{\infty}=
V^{(\mathtt{A})}_{\mathtt{eff}}\Big|_{\infty}=\left[U^{(\mathtt{A})}+V^{(\mathtt{A})}\right]_{\infty},
\end{equation}
where the effective potential $V^{(\mathtt{A})}_{\mathtt{eff}}$
corresponds to Eq. (\ref{veff}) in scaled units.

\subsection{$N$-electron atom infinite-$D$ analysis}\label{sec:atom1}
We have chosen a particularly simple infinite-$D$ configuration:
the totally symmetric configuration, which is characterized by the
equality of all $\bar{r}_i$ and $\gamma_{ij}$. The effective
potential is an extremum when its derivatives with respect to all
$\bar{r}_i$ and $\gamma_{ij}$ are zero at $\bar{r}_{\infty}$ and
$\gamma_{\infty}$. Using the conditions imposed in Eqs.
(\ref{minimum1}) and (\ref{minimum2}), we find:
\begin{eqnarray}
&&-\frac{1}{\bar{r}_{\infty}^3}\frac{1+(N-2)\gamma_{\infty}}
 {(1+(N-1)\gamma_{\infty})(1-\gamma_{\infty})}+\frac{1}{\bar{r}_{\infty}^2}
 \left[ 1-\frac{N-1}{2^{3/2} Z \sqrt{1-\gamma_{\infty}}}\right] =
 0,
 \nonumber\\ &&\frac{1}{\bar{r}_{\infty}^2}\frac{(2+(N-2)
 \gamma_{\infty}) \gamma_{\infty}}
 {(1+(N-1)\gamma_{\infty})^2 (1-\gamma_{\infty})^2}+
 \frac{1}{\bar{r}_{\infty}}\frac{1}{2^{3/2} Z (1-\gamma_{\infty})^{3/2}} =
 0,
\end{eqnarray}
where we have used the infinite-$D$ symmetric-minimum Gramian
results given in Eqs. (\ref{firstgammas}) of Appendix
\ref{app:gram}.

One can eliminate $\bar{r}_{\infty}$ from the above equations and
then $\gamma_{\infty}$ is the negative solution of smallest
magnitude of the following equation:
\begin{equation}\label{eq:tss_sol}
8Z^2 \gamma_{\infty}^2[2+(N-2)\gamma_{\infty}]^2
+\gamma_{\infty}-1 =0.
\end{equation}
For $Z=N$ it has been verified that this extremum is a minimum. If
we consider $Z$ to be a parameter and we decrease this parameter,
the system becomes more and more unstable and at some point the
above extremum is no longer a minimum. This critical value
$Z_{local}$ can be obtained via Eq. (\ref{eq:tss_sol}) from the
negative real solution of the following equation:
\begin{equation}
(4-6\gamma_{\infty})(1-\gamma_{\infty})^3+N\gamma_{\infty}(22-9\gamma_{\infty})
(1-\gamma_{\infty})^2+20
N^2\gamma_{\infty}^2(1-\gamma_{\infty})+5N^3 \gamma_{\infty}^3=0,
\end{equation}
which is simply the condition under which the smallest eigenvalue
of the Hessian\cite{strang} changes sign. $Z_{local}$ is always
less than $N$ for neutral atoms, and so the totally symmetric
state is indeed in an energetic (local) minimum for neutral atoms.

The infinite-$D$ radius and energy can be written in terms of the
solution $\gamma_{\infty}$ of the quartic equation
(\ref{eq:tss_sol}) as follows:
\begin{equation}
\bar{r}^{(\mathtt{A})}_{\infty} = [1+(N-1)\gamma_{\infty}]^{-2}
\end{equation}
\begin{equation}\label{EAinf}
\frac{\bar{E}^{(\mathtt{A})}_{\infty}}{N}=-\frac{1}{2}[1+(N-1)\gamma_{\infty}]^3
[1+(N-2)\gamma_{\infty}]/(1-\gamma_{\infty}).
\end{equation}
We should remark here on the fact that we are considering a
minimum in the effective potential, which allows us to eliminate
$Z$ from the above expressions. Specifically, we have employed the
conditions (\ref{minimum1}) and (\ref{minimum2}) at the minimum
that
\begin{equation}
\frac{\partial V^{(\mathtt{A})}}{\partial
\gamma_{ij}}\Big|_{\infty}= -\frac{\partial
U^{(\mathtt{A})}}{\partial \gamma_{ij}}\Big|_{\infty},
\end{equation}
which leads to
\begin{equation}\label{atomcoupling}
\frac{1}{Z}=-\frac{2^{3/2}\gamma_{\infty}
[2+(N-2)\gamma_{\infty}]}{\sqrt{1-\gamma_{\infty}}}.
\end{equation}
\subsection{$N$-electron atom normal modes}

We can determine the elements of ${\bf G}$ by comparing the
differential terms in Eq. (\ref{Gham}) with  ${\mathcal
T}^{(\mathtt{A})}$ in Eq. (\ref{eq:resc_hamilt}) expanded to first
order in $1/D$ as in Eq. (\ref{eq:T}). The non-zero elements of
the ${\bf G}$ matrix are:
\begin{equation}\label{eq:atomicG}
\begin{array}{lll}
{G}_{a}=1& &\nonumber\\
{G}_{g}=2\frac{1-{\gamma_{\infty}}^2}{\left({\bar{r}^{(\mathtt{A})}_{\infty}}\right)^2}
&=&2[1+(N-1)\gamma_{\infty}]^4(1+\gamma_{\infty})(1-\gamma_{\infty})
\nonumber\\ {G}_{h} =
\frac{\gamma_{\infty}(1-\gamma_{\infty})}{\left({\bar{r}^{(\mathtt{A})}_{\infty}}\right)^2}
&=&[1+(N-1)\gamma_{\infty}]^4\gamma_{\infty}(1-\gamma_{\infty}),
\nonumber
\end{array}
\end{equation}
where the matrix elements have been evaluated at the infinite-$D$
symmetric minimum and we have used the notation in Eq.
(\ref{eq:GFsyma}). The factor of $2$ in $G_g$ comes from the fact
that each $(ij,ij)$ term appears twice in ${\mathcal
T}^{(\mathtt{A})}$ of Eq. (\ref{eq:resc_hamilt}) (e.g., $(21,21)$
as well as $(12,12)$), but it is only counted once in the {\bf G}
matrix (e.g., only $(12,12)$ is counted).  In a similar manner to
the {\bf G}-matrix elements, the non-zero ${\bf F}$-matrix
elements are:
\begin{eqnarray}
{F}_{a}&=&\frac{[1+(N-1)\gamma_{\infty}]^6}{(1-\gamma_{\infty})^2}
\left( \frac{2+2(5N-7)\gamma_{\infty}+(5N-6)(N-3)\gamma_{\infty}^2
-3(N-1)(N-2)\gamma_{\infty}^3}{2}\right)\nonumber\\
{F}_{b}&=&\frac{[1+(N-1)\gamma_{\infty}]^6}{(1-\gamma_{\infty})^2}
\left(\frac{-\gamma_{\infty}(3-\gamma_{\infty})[2+(N-2)\gamma_{\infty}]}{2}
\right)\nonumber\\
{F}_{e}&=&\frac{[1+(N-1)\gamma_{\infty}]^4}{(1-\gamma_{\infty})^2}
\left(
\frac{-\gamma_{\infty}[2+3(N-2)\gamma_{\infty}]}{2}\right)\nonumber\\
{F}_{f}&=&\frac{[1+(N-1)\gamma_{\infty}]^4}{(1-\gamma_{\infty})^2}
\left( 2\gamma_{\infty}^2 \right)\\
{F}_{g}&=&\frac{[1+(N-1)\gamma_{\infty}]}{(1-\gamma_{\infty})^3}
\left(\frac{4+6(2N-5)\gamma_{\infty}+(12N^2-53N+64)\gamma_{\infty}^2
+(4N^3-23N^2+49N-38)\gamma_{\infty}^3}{2}\right)\nonumber\\
{F}_{h}&=&-\frac{[1+(N-1)\gamma_{\infty}]}{(1-\gamma_{\infty})^3}
\gamma_{\infty}\left(3+(5N-14)\gamma_{\infty}+(2N^2-9N+11)\gamma_{\infty}^2\right)
\nonumber\\
{F}_{\iota}&=&\frac{[1+(N-1)\gamma_{\infty}]}{(1-\gamma_{\infty})^3}
4\gamma_{\infty}^2\left(2+(N-2)\gamma_{\infty}\right),\nonumber
\end{eqnarray}
where we have used Eqs. (\ref{secondgammas}) from Appendix
\ref{app:gram} to evaluate the Gramian second-order derivatives.
Again we have used the fact that we are considering a minimum in
the effective potential, which allows us to use Eq.
(\ref{atomcoupling}) to eliminate $Z$ from the above expressions.
Pulling out the common term among the ${\bf GF}$ matrix elements,
we define the matrix $({\bf GF})'$ as:
\begin{equation}\label{eq:H_def}
({\bf GF})'\equiv \tau {\bf G}{\bf F},
\end{equation}
where
\begin{equation}\label{eq:tau}
\tau\equiv
\frac{2(1-\gamma_{\infty})^2}{[1+(N-1)\gamma_{\infty}]^6} \Omega.
\end{equation}
Using the above equations along with Eqs. (\ref{GFsym}) we find
the following for the elements of $({\bf GF})'$:
\begin{eqnarray}\label{eq:gfprime}
{a}&=&2+2(5N-7)\gamma_{\infty}+(5N-6)(N-3)\gamma_{\infty}^2-3(N-1)(N-2)
\gamma_{\infty}^3\nonumber\\
{b}&=&-\gamma_{\infty}(3-\gamma_{\infty})[2+(N-2)\gamma_{\infty}]
\nonumber\\
{c}&=&-\gamma_{\infty}(1-\gamma_{\infty})[2+(3N-4)\gamma_{\infty}]
[2+(N-2)\gamma_{\infty}][1+(N-1)\gamma_{\infty}]^2\nonumber\\
{d}&=&2\gamma_{\infty}^2(1-\gamma_{\infty})[2+(N-2)\gamma_{\infty}]
[1+(N-1)\gamma_{\infty}]^2\nonumber\\
{e}&=&-\gamma_{\infty}[2+3(N-2)\gamma_{\infty}]
[1+(N-1)\gamma_{\infty}]^{-2}\\
{f}&=&4\gamma_{\infty}^2[1+(N-1)\gamma_{\infty}]^{-2} \nonumber\\
{g}&=&2[4+2(4N-11)\gamma_{\infty}+(4N^2-17N+24)\gamma_{\infty}^2+
3(N-2)\gamma_{\infty}^3]\nonumber\\
{h}&=&-\gamma_{\infty}[8+6(N-3)\gamma_{\infty}-(3N-10)
\gamma_{\infty}^2]\nonumber\\
{\iota}&=&8\gamma_{\infty}^2(1-\gamma_{\infty}).\nonumber
\end{eqnarray}
The eigenvalues of $({\bf GF})'$ will be equal to the eigenvalues
of ${\bf GF}$ multiplied by the common factor $\tau$ in Eq.
(\ref{eq:tau}).

Using the matrix elements of Eq. (\ref{eq:gfprime}), we obtain the
two multiplicity-$1$ modes from Eqs. (\ref{eq:charpol0}) and
(\ref{lam0defs}):
\newcommand{\gi}{\gamma_\infty}
\begin{equation}\label{eq:lamA0}
\lambda^{(\mathtt{A})}_{0^{\pm}} =
5(1-\gi)^2+2N\gi(1-\gi)+N^2\gi^2\pm
\sqrt{\begin{array}{r} 3(1-\gi)^4(3+8\gi)-12N\gi(1-\gi)^3(1-3\gi)-\\
2N^2\gi^2(1-\gi)^2(1-6\gi) +4N^3\gi^3(1-\gi)+N^4\gi^4
\end{array}
}.
\end{equation}
For the modes with multiplicity $N-1$
($\lambda^{(\mathtt{A})}_{1^{\pm}}$) and multiplicity $N(N-3)/2$
($\lambda^{(\mathtt{A})}_2$) we find from Eqs. (\ref{eq:charpol2}),
(\ref{eq:charpol1}), and (\ref{lam1defs}):
\begin{equation}\label{eq:lamA1}
\lambda^{(\mathtt{A})}_{1^{\pm}} =
5(1-\gi)^2+9N\gi(1-\gi)+\frac{7}{2}N^2\gi^2\pm
\sqrt{\begin{array}{r} 3(1-\gi)^4(3+8\gi)-6N\gi(1-\gi)^3(1-8\gi)-\\
2N^2\gi^2(1-\gi)^2(4-9\gi) +3N^3\gi^3(1-\gi)+\frac{9}{4}N^4\gi^4
\end{array}},
\end{equation}
\begin{equation}\label{eq:lamA2}
\lambda^{(\mathtt{A})}_2=4(1-\gi)(2-5\gi)+2N\gi(8-11\gi)+8N^2\gi^2.
\end{equation}
With $\tau$ given in Eq. (\ref{eq:tau}), from Eq.
(\ref{eq:omega_p}) the normal-mode frequencies $\omega$ are
related to these $\lambda$'s by
\begin{equation}\label{eq:omA}
\bar{\omega}^{(\mathtt{A})}=\sqrt{\frac{\lambda^{(\mathtt{A})}}{\tau}}.
\end{equation}
Having obtained the normal-mode frequencies from Eqs.
(\ref{eq:lamA0}), (\ref{eq:lamA1}), (\ref{eq:lamA2}), and
(\ref{eq:omA}), the energy through first order is given by Eqs.
(\ref{eq:E1}) and (\ref{EAinf}).

\section{$N$-electron Quantum Dot}
\label{sec:qd} We follow the $N$-electron atom with an analogous
many-electron system, the quantum dot, or, as it is sometimes
called, the artificial atom\cite{ashoori}. Quantum dots are
nanostructures in which a controllable number of electrons are
attracted to a central location.  But instead of the Coulomb
attraction of a nucleus, the quantum dot electrons are attracted
to the center of an external trapping potential. Another
difference between quantum dots and atoms is their size: quantum
dots are typically much larger than atoms.  Furthermore, the
quantum dot electrons typically interact in some medium such as a
semiconductor; thus, we use the effective-mass approximation where
the electrons, each with mass $m^*$, move in a medium with
dielectric constant $\epsilon$. McKinney and Watson have used
dimensional perturbation theory to solve for the two-electron
quantum dot spectrum\cite{qdot}.  In their paper, they outline how
one could use the many-body techniques previously introduced by
Loeser\cite{loeser,loeser2} and detailed in this paper, to apply
dimensional perturbation theory to a quantum dot with an arbitrary
number of electrons.  We now provide the details sketched in Ref.
\cite{qdot} .

The electrons, each of effective mass $m^*$, are confined by a
spherical harmonic trap with trapping frequency
$\omega_{\mathtt{ho}}$:
\begin{equation}
V_{\mathtt{conf}}(r_i)=\frac{1}{2}m^{*}\omega_{\mathtt{ho}}^2{r_i}^2,
\end{equation}
and we take the interelectron potential to be
\begin{equation}
V_{\mathtt{int}}(r_{ij})=\frac{e^2}{\epsilon \sqrt{r_i^2 + r_j^2 -
2 r_i r_j \gamma_{ij}}},
\end{equation}
where $e$ is the electric charge and $\epsilon$ is the dielectric
constant.  Unlike the $N$-electron atom problem, where we used
dimensionally scaled atomic units, for the quantum dot we use
dimensionally scaled harmonic oscillator units. We can transfer
all of the explicit dimension dependence to the differ
ential part
of the kinetic energy and regularize the large-dimension limit by
substituting the following dimensionally scaled variables into the
similarity transformed Schr\"odinger equation (\ref{eq:SE}):
\begin{equation}\label{unitsQD}
\bar{r}_i =\frac{r_i}{\Omega {l}_{\mathtt{ho}}}, \;\;\;\;\;
\bar{E} = \frac{\Omega}{\hbar \bar{\omega}_{\mathtt{ho}}} E, \;\;
\mbox{and} \;\; \Omega=(D-1)(D-2N-1)/4,
\end{equation}
where the dimensionally scaled harmonic oscillator length and trap
frequency are, respectively,
\begin{equation}
{l}_{\mathtt{ho}}=\sqrt{\frac{\hbar}{m^*\bar{\omega}_{\mathtt{ho}}}}\;\;
\mbox{and} \;\; \bar{\omega}^2_{\mathtt{ho}}=\Omega^3
{\omega}^2_{\mathtt{ho}}.
\end{equation}
That is, we obtain
\begin{equation}
({\mathcal
T}^{(\mathtt{QD})}+U^{(\mathtt{QD})}+V^{(\mathtt{QD})})\Phi =
\bar{E} \Phi,
\end{equation}
where
\begin{eqnarray}\label{eq:resc_hamiltQD}
&& {\mathcal
T}^{(\mathtt{QD})}=-\frac{1}{2\Omega}\sum\limits_{i=1}^N \left[
\frac{\partial^2}{{\partial\bar{r}_i}^2}+
\sum\limits_{j\not=i}\sum\limits_{k\not=i}\frac{\gamma_{jk}-\gamma_{ij}
\gamma_{ik}}{{\bar{r}_i}^2}\frac{\partial^2}{\partial\gamma_{ij}\partial\gamma_{ik}}
\right] \nonumber\\
&&U^{(\mathtt{QD})}=\frac{1}{2}\sum\limits_{i=1}^{N}\frac{1}{{\bar{r}_i}^2}\frac{\Gamma^{(i)}}{
\Gamma} \\
&&V^{(\mathtt{QD})}=\sum\limits_{i=1}^{N}\frac{1}{2}\bar{r}_i^2+\frac{1}{\xi}
\sum\limits_{i=1}^{N-1}\sum\limits_{j=i+1}^{N}\frac{1}{\sqrt{{\bar{r}_i}^2+
{\bar{r}_j}^2-2\bar{r}_i\bar{r}_j\gamma_{ij}}},\nonumber
\end{eqnarray}
and where
\begin{equation}\label{eq:xi}
\xi=\frac{a^*}{l_{\mathtt{ho}}} \;\; \mbox{and} \;\;
a^*=\frac{\epsilon \hbar^2}{m^* e^2}
\end{equation}
are the coupling constant and the effective bohr radius,
respectively. From Eq. (\ref{eq:resc_hamiltQD}), it can be seen
that if $D$, and hence $\Omega$, becomes infinitely large, the
differential part ${\mathcal T}^{(\mathtt{QD})}$ of the kinetic
energy will drop out of the Hamiltonian, just as for the
$N$-electron atom. The particles behave as though they become
infinitely heavy and the infinite-D energy becomes the value of
the effective potential,
\begin{equation}\label{eq:veff_qd}
V_{\mathtt{eff}}^{(\mathtt{QD})} =
U^{(\mathtt{QD})}+V^{(\mathtt{QD})},
\end{equation}
at its minimum:
\begin{equation}
\bar{E}^{(\mathtt{QD})}_{\infty}=V_{\mathtt{eff}}^{(\mathtt{QD})}\Big|_{\infty}
= \left[U^{(\mathtt{QD})}+V^{(\mathtt{QD})}\right]_{\infty}.
\end{equation}

\subsection{$N$-electron Quantum Dot infinite-$D$ analysis}
As before, we choose the totally symmetric configuration for which
all $\bar{r}_i$ and $\gamma_{ij}$ are equal to some
$\bar{r}_{\infty}$ and $\gamma_{\infty}$. The effective potential
is an extremum when its derivatives with respect to all
$\bar{r}_i$ and $\gamma_{ij}$ are zero at this $\bar{r}_{\infty}$
and $\gamma_{\infty}$. Using Eq. (\ref{eq:veff_qd}) in Eqs.
(\ref{minimum1}) and (\ref{minimum2}), we find:
\begin{eqnarray}
&&-\frac{1}{\bar{r}_{\infty}^3}\frac{1+(N-2)\gamma_{\infty}}
 {(1+(N-1)\gamma_{\infty})(1-\gamma_{\infty})}+
 \left[{\bar{r}_{\infty}}-\frac{N-1}{2^{3/2} \xi {\bar{r}_{\infty}^2} \sqrt{1-\gamma_{\infty}}}\right] =
 0,
 \\ &&\frac{1}{\bar{r}_{\infty}^2}\frac{(2+(N-2)
 \gamma_{\infty}) \gamma_{\infty}}
 {(1+(N-1)\gamma_{\infty})^2 (1-\gamma_{\infty})^2}+
 \frac{1}{\bar{r}_{\infty}}\frac{1}{2^{3/2} \xi (1-\gamma_{\infty})^{3/2}} =
 0,
\end{eqnarray}
where we have used the infinite-$D$ symmetric minimum Gramian
results given in Eqs. (\ref{firstgammas}) of Appendix
\ref{app:gram}.

One can eliminate $\bar{r}_{\infty}$ from the above equations and
then, just as for the $N$-electron atom, $\gamma_{\infty}$ is the
negative solution of smallest magnitude of a quartic equation:
\begin{equation}\label{eq:tss_solqd}
8\xi^2 \gamma_{\infty}^2[2+(N-2)\gamma_{\infty}]^2
-(1-\gamma_{\infty})(1+(N-1)\gamma_{\infty})^3 =0.
\end{equation}
The infinite-$D$ radius and energy per atom are then given by
\begin{eqnarray}
&& \bar{r}^{(\mathtt{QD})}_{\infty} = [1+(N-1)\gamma_{\infty}]^{-1/2}\\
&& \frac{\bar{E}_{\infty}^{(\mathtt{QD})}}{N}=
-\frac{1}{2}\frac{(N-1)(N-2)\gamma_{\infty}^2+2N\gamma_{\infty}-2}
{(1-\gamma_{\infty})[1+(N-1)\gamma_{\infty}]},\label{QDzeroth}
\end{eqnarray}
where we use the superscript (QD) to distinguish similar
quantities in the atomic and hard-sphere systems.  Just as in the
$N$-electron atom analysis, we have used the fact that we are
considering a minimum in the effective potential, which allows us
to eliminate the coupling constant $\xi$ from the above
expressions. Specifically, we have employed the conditions
(\ref{minimum1}) and (\ref{minimum2}) at the minimum that
\begin{equation}
\frac{\partial V^{(\mathtt{QD})}}{\partial
\gamma_{ij}}\Big|_{\infty}= -\frac{\partial
U^{(\mathtt{QD})}}{\partial \gamma_{ij}}\Big|_{\infty},
\end{equation}
which leads to
\begin{equation}\label{QDcoupling}
\xi=-\frac{2^{3/2}\gamma_{\infty}
[2+(N-2)\gamma_{\infty}]}{\sqrt{1-\gamma_{\infty}}
[1+(N-1)\gamma_{\infty}]^{3/2}}.
\end{equation}

\subsection{$N$-electron Quantum Dot Normal Modes}
We can determine the elements of ${\bf G}$ by comparing the
differential term in Eq. (\ref{Gham}) with  ${\mathcal
T}^{(\mathtt{QD})}$ in Eq. (\ref{eq:resc_hamiltQD}) expanded to
first order in $1/D$ as in Eq. (\ref{eq:T}). The non-zero elements
of the ${\bf G}$ matrix are:
\begin{equation}
\begin{array}{lll}
{G}_{a}=1 &&\nonumber\\
{G}_{g}=2\frac{1-{\gamma_{\infty}}^2}{\left({\bar{r}^{(\mathtt{QD})}_{\infty}}\right)^2}
&=&2(1-\gamma_{\infty})(1+\gamma_{\infty})[1+(N-1)\gamma_{\infty}]
\nonumber\\ {G}_{h}=
\frac{\gamma_{\infty}(1-\gamma_{\infty})}{\left({\bar{r}^{(\mathtt{QD})}_{\infty}}\right)^2}
&=&\gamma_{\infty}(1-\gamma_{\infty})[1+(N-1)\gamma_{\infty}],\nonumber
\end{array}
\end{equation}
where the matrix elements have been evaluated at the infinite-$D$
symmetric minimum and we have used the notation in Eq.
(\ref{eq:GFsyma}). (See the discussion after Eq.
(\ref{eq:atomicG}) for an explanation of the factor of $2$ in
$G_g$.) Likewise, the non-zero ${\bf F}$ matrix elements are:
\begin{eqnarray}
{F}_{a}&=&\frac{1}{2(1-\gamma_{\infty})^2} \left(
8+2(5N-13)\gamma_{\infty}+(5N^2-21N+24)\gamma_{\infty}^2
-3(N-1)(N-2)\gamma_{\infty}^3\right)\nonumber\\
{F}_{b}&=&\frac{\gamma_{\infty}(\gamma_{\infty}-3)(2+(N-2)\gamma_{\infty})}
{2(1-\gamma_{\infty})^2} \nonumber\\
{F}_{e}&=&\frac{-\gamma_{\infty}}
{2(1-\gamma_{\infty})^2(1+(N-1)\gamma_{\infty})^{1/2}}
\left(2+3(N-2)\gamma_{\infty}\right)\nonumber\\
{F}_{f}&=&\frac{2{\gamma_{\infty}}^2}
{(1-\gamma_{\infty})^2(1+(N-1)\gamma_{\infty})^{1/2}}\\
{F}_{g}&=&\frac{1}
{2(1-\gamma_{\infty})^3(1+(N-1)\gamma_{\infty})^{2}}
\Bigl(4+6(2N-5)\gamma_{\infty} + (64-53N+12N^2){\gamma_{\infty}}^2
+ \nonumber \\ &\; \;& (-38+49N-23N^2+4N^3){\gamma_{\infty}}^3\Bigr)\nonumber\\
{F}_{h}&=&\frac{-2\gamma_{\infty}}
{(1-\gamma_{\infty})^3(1+(N-1)\gamma_{\infty})^{2}}
\left(3+(5N-14)\gamma_{\infty} + (11-9N+2N^2){\gamma_{\infty}}^2
\right)\nonumber\\
F_{\iota}&=&
\frac{[1+(N-1)\gamma_{\infty}]}{(1-\gamma_{\infty})^3}
4\gamma_{\infty}^2\left(2+(N-2)\gamma_{\infty}\right), \nonumber
\end{eqnarray}
where we have used Eqs. (\ref{secondgammas}) from Appendix
\ref{app:gram} to evaluate the Gramian second-order derivatives.
Again, since we are considering a minimum in the effective
potential, we employed Eq. (\ref{QDcoupling}) to replace the
coupling constant $\xi$ of Eq. (\ref{eq:xi}) in favor of
$\gamma_{\infty}$. Using the above equations for ${\bf F}$ and
${\bf G}$ along with Eqs. (\ref{GFsym}) we find:
\begin{eqnarray}
{a}&=&\frac{1}{2(1-\gamma_{\infty})^2} \left(
8+2(5N-13)\gamma_{\infty}+(5N^2-21N+24)\gamma_{\infty}^2
-3(N-1)(N-2)\gamma_{\infty}^3\right)\nonumber\\
{b}&=&\frac{\gamma_{\infty}(\gamma_{\infty}-3)(2+(N-2)\gamma_{\infty})}
{2(1-\gamma_{\infty})^2}\nonumber\\
{c}&=&\frac{{-\gamma_{\infty}}(1+\gamma_{\infty})(1+(N-1)\gamma_{\infty})^{1/2}}
{2(1-\gamma_{\infty})} \left(4 + 3(N-2)^2{\gamma_{\infty}}^2
\right)\nonumber\\
{d}&=&\frac{{\gamma_{\infty}}^2 (1+\gamma_{\infty})
(1+(N-1)\gamma_{\infty})^{1/2}}
{(1-\gamma_{\infty})} \left(2+(N-6)\gamma_{\infty}\right)\nonumber\\
{e}&=&\frac{-\gamma_{\infty}}
{2(1-\gamma_{\infty})^2(1+(N-1)\gamma_{\infty})^{1/2}}
\left(2+3(N-2)\gamma_{\infty}\right)\\
{f}&=&\frac{2{\gamma_{\infty}}^2}
{(1-\gamma_{\infty})^2(1+(N-1)\gamma_{\infty})^{1/2}}\nonumber \\
{g}&=&\frac{1} {(1-\gamma_{\infty})^2(1+(N-1)\gamma_{\infty})}
\Bigl(4+2(6N-13)\gamma_{\infty} +
(50-37N+2N^2+2N^3){\gamma_{\infty}}^2+ \nonumber \\
&\; \;&(-30+20N+27N^2-12N^3+2N^4){\gamma_{\infty}}^3 +
(50-107N+89N^2-22N^3+2N^4){\gamma_{\infty}}^4\Bigr)\nonumber\\
{h}&=&\frac{\gamma_{\infty}}
{2(1-\gamma_{\infty})^2(1+(N-1)\gamma_{\infty})}
\Bigl(4+6(2N-7)\gamma_{\infty} +
(16-17N+2N^2+2N^3){\gamma_{\infty}}^2+ \nonumber \\
&\; \;&(-14-35N+45N^2-14N^3+2N^4){\gamma_{\infty}}^3\Bigr)\nonumber\\
{\iota}&=&\frac{-4{\gamma_{\infty}}^2 (1+\gamma_{\infty})}
{(1-\gamma_{\infty})^2(1+(N-1)\gamma_{\infty})}
\left(-1-(N-6)\gamma_{\infty} +
(3N-5){\gamma_{\infty}}^2\right).\nonumber
\end{eqnarray}

Substituting these ${\bf GF}$ matrix elements in Eqs.
(\ref{eq:omega_p}), (\ref{eq:charpol2}), (\ref{eq:charpol1}), and
(\ref{eq:charpol0}), we now write down the quantum dot normal-mode
frequencies. The multiplicity-$N(N-3)/2$ mode is a vector mode
with frequency
\begin{equation}\label{QDom2}
\bar{\omega}_2^{(\mathtt{QD})}=\sqrt{g-2h+\iota}.
\end{equation}
The multiplicity-$(N-1)$ asymmetric stretch and bend frequencies
take the form
\begin{equation}\label{QDom1}
\bar{\omega}_{{1}^{\pm}}^{(\mathtt{QD})}=\sqrt{\eta_1 \pm
\sqrt{{\eta_1}^2-\Delta_1}},
\end{equation}
and the two multiplicity-$1$ symmetric stretch and bend
frequencies take the form
\begin{equation}\label{QDom0}
\bar{\omega}_{{0}^{\pm}}^{(\mathtt{QD})}=\sqrt{\eta_0 \pm
\sqrt{{\eta_0}^2-\Delta_0}},
\end{equation}
where $\eta_1$ and $\Delta_1$ are given in Eq. (\ref{lam1defs}),
and $\eta_0$ and $\Delta_0$ are given in Eq. (\ref{lam0defs}).

Having obtained the frequencies from Eqs. (\ref{QDom2}),
(\ref{QDom1}), and (\ref{QDom0}) and the Lewis structure energy in
Eq. (\ref{QDzeroth}), the energy through first order is given by
Eq. (\ref{eq:E1}). Although this is only a low order
approximation, two-body studies suggest that when the coupling
constant $\xi<<1$ (i.e., the strongly interacting regime where the
repulsive energy of the electrons dominates the confinement energy
of the trap) low order dimensional perturbation may be very
accurate\cite{qdot}. Dimensional perturbation theory has a
non-perturbative character in the sense that the leading-order
term of DPT includes a contribution from the Coulomb potential.

\section{$N$ hard-sphere particles in a trap: Atomic Bose-Einstein condensates}
\label{sec:hs} In this section we treat $N$ hard spheres in a
trap. An important application of this many-body system is $N$
trapped ultra-cold Bose atoms (i.e., an inhomogeneous
Bose-Einstein condensate)\cite{becreview}. Dimensional
perturbation techniques were introduced to inhomogeneous atomic
BEC in Ref.\cite{dptgp} by scaling the non-linear mean-field
(Gross-Pitaevskii) equation. The treatment in this section goes
beyond the mean-field approximation as it includes correlation and
uses a finite-range, as opposed to a zero-range, interatomic
potential.

The atoms of equal mass $m$ are confined by a spherical harmonic
trap with trapping frequency $\omega_{\mathtt{ho}}$:
\begin{equation}
V_{\mathtt{conf}}(r_i)=\frac{1}{2}m\omega_{\mathtt{ho}}^2{r_i}^2.
\end{equation}
We take the interaction potential to be a hard sphere of radius
$a$ (also the scattering length for the BEC system):
\begin{equation}
V_{\mathtt{int}}(r_{ij})=
\left\{ \begin{array}{ll} \infty, & r_{ij} < a \\
0,& r_{ij} \ge a.
\end{array}\right.
\end{equation}
In the two previous systems, many-electron systems, the
dimensionally continued Laplacian is dimension dependent while the
potential energy maintains the same form as it has at $D=3$.  In
this system, we dimensionally continue the interaction potential
so that in the large-$D$ limit the effective potential is
differentiable, and thus allowing us to perform the dimensional
perturbation analysis outlined in Secs. \ref{sec:infD} and
\ref{sec:firstorder} which requires taking first- and second-order
derivatives of the effective potential. Specifically, we take the
interaction to be
\begin{equation}
V_{\mathtt{int}}(r_{ij})=\frac{V_o}{1-3/D}\left[1-\tanh\left\{\frac{c_o}{1-3/D}
\left(\frac{r_{ij}}{\sqrt{2}}-\alpha-\frac{3}{D}(a-\alpha)\right)\left(1+(1-3/D)
\sum_{n=1}^{s-3}\frac{c_n r_{ij}^{2n}}{2^n}\right) \right\}
\right],
\end{equation}
where $D$ is the Cartesian dimensionality of space.  This
interaction becomes a hard-sphere of radius $a$ in the physical,
$D=3$, limit. The other $s$ constants ($V_o$, $\alpha$, and
$\{c_n; \mbox{ } \forall \mbox{ } n: 0 \le n \le s-3\}$) are parameters
that allow us to fine-tune the large-$D$ shape of the potential and optimize
our results through Langmuir (first) order\cite{mbbecpaper}. The
simplest possibility could have as few as two parameters: $V_o$
and $c_o$, with $\alpha=a$ and the remaining $c_n=0$; however, we
can have any number of parameters for the most general and
flexible potential.  To see how the potential may be fine-tuned to
optimize the results through Langmuir order, see Ref.
\cite{mbbecpaper}.

We use dimensionally scaled harmonic oscillator units similar to
the $N$-electron quantum dot system with the difference that we
use $D^2$ in the dimensional scaling instead of $\Omega$ of Eq.
(\ref{unitsQD}). Ideally, one would use $\Omega$ in order to
remove all of the dimension dependence from the centrifugal-like
term in the kinetic energy, but in order to simplify the scaling
of our dimensionally continued hard-sphere potential, we allow
some dimension dependence in the centrifugal-like term (see
$U^{(\mathtt{HS})}$ in (\ref{eq:resc_hamiltHS}) below). We
regularize the large-$D$ limit of the Schr\"odinger equation by
using the following dimensionally scaled variables:
\begin{equation}\label{unitsHS}
\begin{array}{l}
\bar{r}_i =\frac{r_i}{D^2 \bar{a}_{\mathtt{ho}}}\;\;\; \bar{E} =
\frac{D^2}{\hbar \bar{\omega}_{\mathtt{ho}}} E, \;\;\;  \bar{H} =
\frac{D^2}{\hbar \bar{\omega}_{\mathtt{ho}}} H, \;\;\; \bar{a}
=\frac{a}{\sqrt{2} D^2 \bar{a}_{\mathtt{ho}}}, \nonumber \\
\bar{V}_{o} = \frac{D^2}{\hbar \bar{\omega}_{\mathtt{ho}}} V_{o},
\;\;\; \bar{\alpha} =\frac{\alpha}{\sqrt{2} D^2 \bar{a}_{\mathtt{ho}}},
\;\;\; \bar{c}_{o}= \sqrt{2} D^2 \bar{a}_{\mathtt{ho}}c_o, \;\;\;
\bar{c}_{n}=(\sqrt{2} D^2 \bar{a}_{\mathtt{ho}})^{2n}c_n,
\end{array}
\end{equation}
where
\begin{equation}\label{unitsHSa}
\bar{a}_{\mathtt{ho}}=\sqrt{\frac{\hbar}{m
\bar{\omega}_{\mathtt{ho}}}} \;\;\; \mbox{and} \;\;\;
{\bar{\omega}_{\mathtt{ho}}}=D^3{\omega_{\mathtt{ho}}}
\end{equation}
are the dimensionally-scaled harmonic-oscillator length scale and
dimensionally-scaled trap frequency, respectively. Substituting
these scaled variables into the similarity transformed
Schr\"odinger equation, Eq. (\ref{eq:SE}), gives the following
equation:
\begin{equation}
\left( {\mathcal
T}^{(\mathtt{HS})}+U^{(\mathtt{HS})}+V^{(\mathtt{HS})} \right)\Phi
= \bar{E} \Phi,
\end{equation}
where
\begin{eqnarray}\label{eq:resc_hamiltHS}
{\mathcal T}^{(\mathtt{HS})}&=&-\frac{1}{2}\delta^2
\sum\limits_{i=1}^N \left[
\frac{\partial^2}{{\partial\bar{r}_i}^2}+
\sum\limits_{j\not=i}\sum\limits_{k\not=i}\frac{\gamma_{jk}-\gamma_{ij}
\gamma_{ik}}{{\bar{r}_i}^2}\frac{\partial^2}{\partial\gamma_{ij}\partial\gamma_{ik}}
\right] \nonumber\\
U^{(\mathtt{HS})}&=&\sum\limits_{i=1}^{N}\frac{(\delta-1)[(2N+1)\delta-1]}{8
{\bar{r}_i}^2}\frac{\Gamma^{(i)}}{
\Gamma} \\
V^{(\mathtt{HS})}&=&\sum\limits_{i=1}^{N}\frac{1}{2}\bar{r}_i^2+ \nonumber\\
&\; \;&+ \frac{\bar{V}_{o}}{1-3\delta}
\sum\limits_{i=1}^{N-1}\sum\limits_{j=i+1}^{N}
\left(1-\tanh\left[\frac{\bar{c}_o}{1-3\delta}
\left(\frac{\bar{r}_{ij}}{\sqrt{2}}-\bar{\alpha}-3\delta\left(
\bar{a}-\bar{\alpha} \right) \right)
\left(1+(1-3\delta)\sum_{n=1}^{s-3} \frac{\bar{c}_n
\bar{r}_{ij}^{2n}}{2^n}\right) \right] \right) \nonumber,
\end{eqnarray}
where $\delta=1/D$ is the perturbation parameter and the
interatomic separation is
\begin{equation}\label{rbarij}
\bar{r}_{ij}={\sqrt{{\bar{r}_i}^2+
{\bar{r}_j}^2-2\bar{r}_i\bar{r}_j\gamma_{ij}}}.
\end{equation}
As $D$ becomes infinitely large, and $\delta\to0$, the entire
differential part of the kinetic energy as well as a portion of
the interatomic and centrifugal-like potentials will drop out of
the Hamiltonian. In the infinite-dimension limit, the particles
behave as though they become infinitely heavy and rest at the
bottom of the infinite-$D$ effective potential, a potential which
includes the confining potential and contributions from the
centrifugal-like and interparticle potentials. The infinite-$D$
energy becomes the minimum value of the effective potential
(\ref{zeroth}).

\subsection{Hard-sphere infinite-$D$ analysis}
The infinite-$D$ ($\delta \to 0$) effective potential in
dimensionally scaled harmonic oscillator units is
\begin{equation}
V_{\mathtt{eff}}^{(\mathtt{HS})}=\sum\limits_{i=1}^{N}\left(
\frac{1}{8 {\bar{r}_i}^2}\frac{\Gamma^{(i)}}{
\Gamma}+\frac{1}{2}\bar{r}_i^2\right)+
\bar{V}_{o}\sum\limits_{i=1}^{N-1}\sum\limits_{j=i+1}^{N}
\left(1-\tanh\left[\bar{c}_{o} \left(\frac{\bar{r}_{ij}}{\sqrt{2}}
-\bar{\alpha}\right) \left(1+\sum_{n=1}^{s-3} \frac{\bar{c}_n
\bar{r}_{ij}^{2n}}{2^n}\right) \right] \right).
\end{equation}
As one can see from the double-sum term in
$V_{\mathtt{eff}}^{(\mathtt{HS})}$, the interaction potential
becomes a soft sphere of radius approximately $\bar{\alpha}$ and
height $2\bar{V}_{o}$.  The slope of the soft wall is determined
by $\bar{c}_{o}$, while, as discussed earlier, the remaining $s-3$
parameters act to further refine the shape of the interaction
potential\cite{mbbecpaper}.

Again choosing the totally symmetric configuration for which all
$\bar{r}_i$ and $\gamma_{ij}$ are equal to some $\bar{r}_{\infty}$
and $\gamma_{\infty}$, and using Eqs. (\ref{firstgammas}) from
Appendix \ref{app:gram} in Eqs. (\ref{minimum1}) and
(\ref{minimum2}), we find that the large-$D$ radii and energy per
atom are:
\begin{eqnarray}
\bar{r}_{\infty}&=& [2(1+(N-1)\gamma_{\infty})]^{-1/2}\\
\frac{\bar{E}_{\infty}^{(\mathtt{HS})}}{N}&=&\frac{1+(N-2)\gamma_{\infty}}{
(1-\gamma_{\infty})(1+(N-1)\gamma_{\infty})}\frac{1}{8{\bar{r}_{\infty}}^2}
+ \frac{1}{2}{\bar{r}_{\infty}}^2 +
\frac{N-1}{2}\bar{V}_{o}\left[1-\tanh(\Theta) \right],
\end{eqnarray}
where for simplicity of presentation we have defined the
following:
\begin{eqnarray}\label{Theta}
\Theta=\bar{c}_{o}\left(\bar{r}_{\infty}\sqrt{1-\gamma_{\infty}}-\bar{\alpha}
\right) \left(1+\sum_{n=1}^{s-3}\bar{c}_{n}\bar{r}_{\infty}^{2n}
(1-\gamma_{\infty})^{n}\right).
\end{eqnarray}

The large-$D$ direction cosine, $\gamma_{\infty}$, of the
hyperangle between the infinite-dimensional radii is given by the
negative solution of smallest magnitude of
\begin{equation}\label{eq:tss_solHS}
\bar{V}_{o}\bar{c}_{o}\Upsilon \mbox{sech}^2\Theta +
\gamma_{\infty}\frac{2(2+(N-2)\gamma_{\infty})^2}{(1-\gamma_{\infty})^3
(1+(N-1)\gamma_{\infty})} =0,
\end{equation}
where
\begin{equation}
\Upsilon=\left[1+\sum_{n=1}^{s-3}\left(
(2n+1)\bar{c}_{n}\bar{r}_{\infty}^{2n}(1-\gamma_{\infty})^{n} - 2n
\bar{\alpha} \bar{c}_{n}
\bar{r}_{\infty}^{2n-1}(1-\gamma_{\infty})^{n-1/2} \right)
\right].
\end{equation}

\subsection{Hard-sphere Normal Modes}

We can determine the elements of ${\bf G}$ by comparing the
differential term in Eq. (\ref{Gham}) with  ${\mathcal
T}^{(\mathtt{HS})}$ in Eq. (\ref{eq:resc_hamiltHS}) expanded to
first order in $1/D$ as in Eq. (\ref{eq:T}). The non-zero elements
of the ${\bf G}$ matrix are found to be:
\begin{equation}
\label{eq:G_HS}
\begin{array}{lll}
{G}_{a}=1 & &\nonumber\\
{G}_{g}=2\frac{1-{\gamma_{\infty}}^2}{{\bar{r}_{\infty}}^2}
&=&4(1-\gamma_{\infty})(1+\gamma_{\infty})[1+(N-1)\gamma_{\infty}]
\nonumber\\ {G}_{h}=
\frac{\gamma_{\infty}(1-\gamma_{\infty})}{{\bar{r}_{\infty}}^2}
&=&2\gamma_{\infty}(1-\gamma_{\infty})[1+(N-1)\gamma_{\infty}],
\nonumber
\end{array}
\end{equation}
where the matrix elements have been evaluated at the infinite-$D$
symmetric minimum and we have used the notation in Eq.
(\ref{eq:GFsyma}).  (See the discussion after Eq.
(\ref{eq:atomicG}) for an explanation of the factor of $2$ in
$G_g$.)  Likewise, the non-zero ${\bf F}$ matrix elements are:
\begin{eqnarray}
\label{eq:F_HS} {F}_{a}&=&1+\frac{3}{4 \bar{r}_{\infty}^4}
\frac{1+(N-2)\gamma_{\infty}}{(1-\gamma_{\infty})(1+(N-1)\gamma_{\infty})}
+\frac{\bar{V}_{o} \bar{c}_{o}}{2}(N-1) \mbox{sech}^2\Theta
\Biggl[\bar{c}_{o}(1-\gamma_{\infty}) \Upsilon^2 \tanh\Theta + \nonumber\\
&\; \;& - \frac{1+\gamma_{\infty}} {2
\bar{r}_{\infty}\sqrt{1-\gamma_{\infty}}} +
\sum_{n=1}^{s-3}\Biggl(\frac{2n+1}{2}\bar{r}_{\infty}^{2n-1}
(1-\gamma_{\infty})^{n-1/2}[(2n-1)\gamma_{\infty}-(2n+1)]+ \nonumber\\
&\; \;& 2n\bar{\alpha}\bar{r}_{\infty}^{2n-2}
(1-\gamma_{\infty})^{n-1}[(n-1)\gamma_{\infty}-n)]
\Biggr)\bar{c}_{n}\Biggr] \nonumber \\
{F}_{b}&=&\frac{\bar{V}_{o} \bar{c}_{o}}{2}\mbox{sech}^2\Theta
\Biggl[\bar{c}_{o}(1-\gamma_{\infty}) \Upsilon^2 \tanh\Theta +
\frac{1+\gamma_{\infty}} {2
\bar{r}_{\infty}\sqrt{1-\gamma_{\infty}}} + \nonumber\\
&\; \;& +
\sum_{n=1}^{s-3}\Biggl(\frac{2n+1}{2}\bar{r}_{\infty}^{2n-1}
(1-\gamma_{\infty})^{n-1/2}[(2n+1)\gamma_{\infty}-(2n-1)]+ \nonumber\\
&\; \;& \;\;\;\;\;\;\;\;\;\;\;\;\;\;\;\;\;\;\;\;\;\;\;\; -
2n\bar{\alpha}\bar{r}_{\infty}^{2n-2}
(1-\gamma_{\infty})^{n-1}[n\gamma_{\infty}-(n-1))]
\Biggr)\bar{c}_{n}\Biggr] \nonumber \\
{F}_{e}&=&-\frac{\gamma_{\infty}}{2 \bar{r}_{\infty}^3}
\frac{1+(N-2)\gamma_{\infty}}{(1-\gamma_{\infty})^2(1+(N-1)\gamma_{\infty})^2}
+\frac{\bar{V}_{o} \bar{c}_{o}}{2} \mbox{sech}^2\Theta
\Biggl[-\bar{c}_{o}\bar{r}_{\infty} \Upsilon^2 \tanh\Theta + \nonumber\\
&\; \;& + \frac{1} {2 \sqrt{1-\gamma_{\infty}}} +
\sum_{n=1}^{s-3}\frac{1}{2} \Biggl((2n+1)^2\bar{r}_{\infty}^{2n}
(1-\gamma_{\infty})^{n-1/2}
-(2n)^2\bar{\alpha}\bar{r}_{\infty}^{2n-1}
(1-\gamma_{\infty})^{n-1}\Biggr)\bar{c}_{n}\Biggr] \nonumber \\
{F}_{f}&=&\frac{{\gamma_{\infty}}^2}
{2\bar{r}_{\infty}^3(1-\gamma_{\infty})^2(1+(N-1)\gamma_{\infty})^{2}}\\
{F}_{g}&=&\frac{1}
{2\bar{r}_{\infty}^2(1-\gamma_{\infty})^3(1+(N-1)\gamma_{\infty})^{3}}
\Bigl(1+3(N-2)\gamma_{\infty} + (13-11N+3N^2){\gamma_{\infty}}^2 +
\nonumber \\ &\; \;& (N-2)(4-3N+N^2){\gamma_{\infty}}^3\Bigr)+
\frac{\bar{V}_{o} \bar{c}_{o}}{2}\mbox{sech}^2\Theta
\Biggl[\frac{\bar{c}_{o}\bar{r}_{\infty}}{1-\gamma_{\infty}}
\Upsilon^2 \tanh\Theta + \frac{\bar{r}_{\infty}}{2
(1-\gamma_{\infty})^{3/2}} + \nonumber\\
&\; \;& -
\sum_{n=1}^{s-3}\Biggl(\frac{(2n-1)(2n+1)}{2}\bar{r}_{\infty}^{2n-1}
(1-\gamma_{\infty})^{n-3/2}+ \nonumber\\
&\; \;& \;\;\;\;\;\;\;\;\;\;\;\;\;\;\;\;\;\;\;\;\;\;\;\; +
2n(n-1)\bar{\alpha}\bar{r}_{\infty}^{2n} (1-\gamma_{\infty})^{n-2}
\Biggr)\bar{c}_{n}\Biggr] \nonumber\\
{F}_{h}&=&\frac{-\gamma_{\infty}}
{4\bar{r}_{\infty}^2(1-\gamma_{\infty})^3(1+(N-1)\gamma_{\infty})^{3}}
\left[3+(5N-14)\gamma_{\infty} + (11-9N+2N^2){\gamma_{\infty}}^2
\right],\nonumber\\
F_{\iota}&=& \frac{\gamma_{\infty}^2
(2+(N-2)\gamma_{\infty})}{\bar{r}_{\infty}^2(1-\gamma_{\infty})^3
(1+(N-1)\gamma_{\infty})^3} \nonumber
\end{eqnarray}
where we have used Eqs. (\ref{secondgammas}) from Appendix
\ref{app:gram} to evaluate the Gramian second derivatives.  For
transparency we leave the $\bar{r}_{\infty}$ terms in Eq.
(\ref{eq:F_HS}) instead of using the explicit $\gamma_{\infty}$
dependent form of $\bar{r}_{\infty}$. The above expressions for
${\bf F}$ and ${\bf G}$ are used to write the elements of ${\bf
GF}$ [i.e., the scalar quantities $a$ through $\iota$ in Eq.
(\ref{GFsym})]. We may then write the five normal-mode frequencies
by means of Eqs. (\ref{eq:charpol2}), (\ref{eq:charpol1}),
(\ref{eq:charpol0}), and (\ref{eq:omega_p}). For the
multiplicity-$N(N-3)/2$ mode, the frequency is given by
\begin{equation}
\bar{\omega}_2^{(\mathtt{HS})}=\sqrt{g-2h+\iota}.
\end{equation}
For the two multiplicity-$(N-1)$ modes, the frequencies are of the
form
\begin{equation}
\bar{\omega}_{{1}^{\pm}}^{(\mathtt{HS})}=\sqrt{\eta_1 \pm
\sqrt{{\eta_1}^2-\Delta_1}},
\end{equation}
and the two multiplicity-$1$ frequencies are of the form
\begin{equation}
\bar{\omega}_{{0}^{\pm}}^{(\mathtt{HS})}=\sqrt{\eta_0 \pm
\sqrt{{\eta_0}^2-\Delta_0}},
\end{equation}
where $\eta_1$ and $\Delta_1$ are given in Eq. (\ref{lam1defs}),
and $\eta_0$ and $\Delta_0$ in Eq. (\ref{lam0defs}).

Because of the factors of $\delta$ in the centrifugal-like and
hard-sphere potentials ($U^{(\mathtt{HS})}$ and
$V^{(\mathtt{HS})}$ of Eq. (\ref{eq:resc_hamiltHS})), there is a
constant shift, $v_{o}$, in the first-order energies:
\begin{equation}\label{eq:ehs}
\overline{E}^{(\mathtt{HS})} = \overline{E}^{(\mathtt{HS})}_{\infty}
+ \delta \Biggl[
\sum_{\renewcommand{\arraystretch}{0}
\begin{array}[t]{r@{}l@{}c@{}l@{}l} \scriptstyle \mu = \{
  & \scriptstyle \mathbf{0}^\pm,\hspace{0.5ex}
  & \scriptstyle \mathbf{1}^\pm & , & \\
  & & \scriptstyle \mathbf{2} & & \scriptstyle  \}
            \end{array}
            \renewcommand{\arraystretch}{1} }
\hspace{-0.50em}
\sum_{\mathsf{n}_{\mu}=0}^\infty ({\mathsf{n}}_{\mu}+\frac{1}{2})
d_{\mu,\mathsf{n}_{\mu}} \bar{\omega}^{(\mathtt{HS})}_{\mu} \,
+ \, v_o \Biggr] \,,
\end{equation}
where the $\mathsf{n}_{\mu}$ are the vibrational quantum numbers
of the normal modes of the same frequency
$\bar{\omega}^{(\mathtt{HS})}_{\mu}$, and
$d_{\mu,\mathsf{n}_{\mu}}$ is the occupancy of the manifold of normal
modes with vibrational quantum number $\mathsf{n}_{\mu}$ and normal
mode frequency $\bar{\omega}^{(\mathtt{HS})}_{\mu}$ (see the
discussion after Eq. (\ref{eq:E1})).  The FG matrix
method gives the normal-mode frequencies of the infinite-$D$
minimum structure. However, due to the explicit
dimension-dependence (i.e., $\delta=1/D$) in $U^{(\mathtt{HS})}$
and $V^{(\mathtt{HS})}$, one must also take into account the
explicit constant factors of order $\delta$ that appear in the
power expansion of Eq. (\ref{eq:resc_hamiltHS}) which are not the
result of second derivatives of the effective potential. This
contribution from the power series is the order-$\delta$ shift,
$v_o$, in the energy in Eq. (\ref{eq:ehs}), and is given by:
\begin{eqnarray}
v_{o}&=&-\frac{N(N+1)(1+(N-2)\gamma_{\infty})}
{4\bar{r}_{\infty}^2(1+(N-1)\gamma_{\infty})(1-\gamma_{\infty})} +
3\bar{V}_{o}\frac{N(N-1)}{2}\Biggl(1-\tanh\Theta+ \nonumber \\ &\;
\;& +
\bar{c}_{o}\left[(\bar{a}-\bar{\alpha})\left(\sum_{n=1}^{s-3}\bar{c}_{n}\bar{r}_{\infty}^{2n}
(1-\gamma_{\infty})\right)-(\bar{r}_{\infty}\sqrt{1-\gamma_{\infty}}-\bar{a})
\right] \mbox{sech}^2\Theta \Biggr),
\end{eqnarray}
where $\Theta$ is given by Eq. (\ref{Theta}).

\section{Summary}


DPT has advantages over traditional methods that either neglect
the interaction at lowest order (conventional perturbation theory)
or the kinetic energy at lowest order (e.g., the large-$N$
Thomas-Fermi approximation\cite{baym,dptgp}). The
infinite-dimension limit (leading order) of DPT results in an
effective potential that keeps contributions from the confinement,
interaction, and kinetic terms of the Hamiltonian. The kinetic
contribution that remains in the large-$D$ limit is a
centrifugal-like ($1/r^2$) term that allows the leading order of
DPT to satisfy the minimum uncertainty principle. Correlation is
included at all orders, including the large-dimension leading
order of the $1/D$ expansion. In the infinite-dimension limit, we
choose a completely symmetric configuration in which all of the
particles are localized and equivalent. The first-order energy
correction corresponds to small oscillations about this high-$D$
symmetric structure, with the vibrational frequencies determined
by the Wilson FG matrix method. There are five distinct roots of
the ${\bf GF}$ matrix, belonging to three different irreducible
representations\cite{loeser,hamermesh} of the symmetric group of
order $N$, $S_N$. These five distinct roots are directly related
to the normal mode frequencies of the many-body system. The two
multiplicity-$(N-1)$ roots, which are a mixture of asymmetric
stretching and bending motions, are designated by ${\bf 1}^-$ and
${\bf 1}^+$, respectively. The two multiplicity-$1$ roots, which
are a mixture of symmetric stretching and bending motions, are
designated by ${\bf 0}^-$ and ${\bf 0}^+$, respectively.  The
multiplicity-$N(N-3)/2$ root, designated by ${\bf 2}$, is a purely
angular mode.

Analytical results through first order are obtained for a
spherical system of identical particles with a general confining
potential and general two-body interaction potential. Many-body
DPT is applied to two $N$-electron systems: the $N$-electron atom
originally studied in Loeser's seminal paper\cite{loeser}
where the confinement is supplied by the attraction of the nucleus,
and the $N$-electron quantum dot where the confinement is supplied
by an external harmonic potential. We also consider the
inhomogeneous bose condensate whose $N$ atoms interact via a
hard-core potential. As in the $N$-electron quantum dot, the
condensate atoms are confined by an external harmonic trap. Unlike
either $N$-electron system, however, the hard sphere potential has
explicit dimension dependence.  Both the atom and quantum dot
interact via a repulsive Coulomb potential that maintains its
three-dimensional form as $D$ varies (an alternative
generalization of the three-dimensional Coulomb potential might be
$1/r^{D-2}$). In order to make the hard sphere amenable to DPT
analysis, we dimensionally continue the hard-sphere potential so
that it is differentiable in the infinite-$D$ limit and becomes a
hard-sphere with radius equal to the scattering length in the
physical $D=3$ limit. This dimensional continuation results in a
soft-sphere interatomic potential at large $D$.

While the treatment of these physical systems has been in many
respects quite general, there remain many avenues for further
development of this many-body DPT formalism. One extension might
be to relax the spherical symmetry constraint by allowing
confining potentials with axial symmetry. This is a particularly
important generalization for describing many experiments,
especially Bose-Einstein condensates, which in practice are
characterized by axial symmetry. With axial symmetry, one has $N$
additional coordinates due to the $z$-components of the $N$
particles, but the many-body formalism presented in this paper
readily accommodates this generalization. Instead of generalized
spherical coordinates with a $D$-dimensional radius, one uses
generalized cylindrical coordinates, which consists of a
$(D-1)$-dimensional sphere plus a $z$-coordinate. Another
extension involves the choice of confining potential.  This paper
focuses mainly on harmonic traps because this is the most
prevalent form of confinement in experiments today, but there is
nothing to prevent one from also applying these methods to more
general external confining potentials. Yet another extension
involves the calculation of higher-order perturbation terms.  The
leading-order DPT wavefunction is a product of one-dimensional
harmonic oscillator wavefunctions with frequencies given in the
first-order term of the energy series (from the FG method). From
this wavefunction, one can then use ordinary perturbation theory
to extend these results to include corrections beyond first order.
This is also an important extension as it should improve DPT's
numerical results for many-particle systems.  Even though the
$1/D$ perturbation series is asymptotic, with appropriate
treatment higher order terms can be used to improve the accuracy
of the low order approximation.

Numerical results are given in Refs. \cite{loeser} and
\cite{mbbecpaper} for the $N$-electron atom and $N$-atom
condensate, respectively. Numerical analysis for a two-electron
quantum dot in Ref. \cite{qdot} suggests that the low-order
analytical many-body results given in Sec. \ref{sec:qd} will be
quite accurate for typical many-electron quantum dots whose length
scales are in the mesoscopic regime (i.e., the strongly
interacting or strongly confined regime). Moreover, because $N$ is
a parameter in our formalism, the challenge of calculating the
physical properties of a given system with DPT does not increase
as one adds more particles. The methods detailed in Secs 2-4 can
be applied to any variety of many-body systems, and we hope that
the systems described in Secs 5-7 are sufficiently varied to
encourage further interest in this novel many-body approach.

\section{Acknowledgments}
This work was supported in part by the Office of Naval Research.
We thank Blake Laing for helpful comments on the manuscript.
\appendix
\renewcommand{\theequation}{A\arabic{equation}}
\setcounter{equation}{0}
\section{Wilson FG matrix method}\label{app:wilson}
In this appendix, we derive the Wilson FG matrix method\cite{dcw}
which is at the heart of our obtaining the normal mode frequencies
and, thereby, the first-order energy correction. The derivation
involves a transformation to the set of coordinates called
normal-coordinates in which both the differential term and the
potential term of Eq. (\ref{Gham}) are diagonal. We begin by
defining a symmetric transformation, ${\bf A}$, that transforms
from the vector $\bar{y}'$, defined by Eq. (\ref{eq:ytransposeP}),
to $\bar{z}'$ ($\bar{z}'={\bf A} \bar{y}'$). ${\bf A}$ is an
active transformation and it has the property that it diagonalizes
the symmetric ${\bf G}$ matrix to unity. That is, ${\bf A}$
satisfies
\begin{eqnarray}
&& {\bf A}^{T}{\bf G}{\bf A}={\bf I},
\end{eqnarray}
which can also be written as
\begin{eqnarray}
\label{A2G} && {\bf G}={\bf A}^{-1}({\bf A}^{-1})^T,
\end{eqnarray}
where we have used the property that ${\bf A}$ is symmetric (${\bf
A}={\bf A}^{T}$) in the derivation of Eq. (\ref{A2G}).
$\widehat{H}_1$ then becomes
\begin{equation}
\label{eq:H1} \widehat{H}_1 \to
-\frac{1}{2}\partial_{\bar{z}'}^{T}\partial_{\bar{z}'}
+\frac{1}{2}\bar{z}'^{T} ({\bf A}^{-1})^{T}{\bf F} {\bf A}^{-1}
\bar{z}'.
\end{equation}

Next we focus our attention on the potential term, the term
involving the matrix ${\bf F}$. We introduce another
transformation, ${\bf U}$ ($\bar{\mathsf q}'={\bf U}\bar{z}'$),
that diagonalizes the potential term while simultaneously leaving
the differential term unchanged. This orthogonal transformation
(${\bf U}^T{\bf U}={\bf I}$, where ${\bf I}$ is the identity
matrix) leaves the differential term in the same form as in Eq.
(\ref{eq:H1}), and the potential term becomes
\begin{equation}\label{tempeig}
{\bf U} ({\bf A}^{-1})^T {\bf F} {\bf A}^{-1} {\bf U}^T = {\bf
\Lambda},
\end{equation}
where $\Lambda$ is a diagonal matrix.  That is,
\begin{equation}\label{eq:appH1}
\widehat{H}_1 \to -\frac{1}{2}\partial_{\bar{\mathsf
q}'}^{T}\partial_{\bar{\mathsf q}'} +\frac{1}{2}\bar{\mathsf
q}'^{T} {\bf \Lambda} \bar{\mathsf q}'.
\end{equation}
The eigenvalue equation corresponding to Eq. (\ref{tempeig}) is
\begin{equation}\label{UA2}
\left({\bf U} ({\bf A}^{-1})^T {\bf F} {\bf A}^{-1} {\bf
U}^T\right) \bar{\mathsf q}'= {\bf \lambda} \bar{\mathsf q}',
\end{equation}
where $\bar{\mathsf q}'$ is the eigenvector with eigenvalue
$\lambda$. Equation (\ref{tempeig}) is a matrix eigenvalue
equation expressed in the basis of the normal coordinates.  We can
change the basis back to the original internal displacement
coordinates with the passive similarity transformation
$\bar{q}'={\bf A}^{-1} {\bf U}^T \bar{\mathsf q}'$, under which
Eq. (\ref{UA2}) now reads
\begin{equation}
{\bf U} ({\bf A}^{-1})^T {\bf F}\bar{q}'= {\bf \lambda} {\bf UA}
\bar{q}'.
\end{equation}
Multiplying on the left by ${\bf U}^T$ followed by ${\bf A^{-1}}$
gives
\begin{equation}
{\bf A}^{-1} ({\bf A}^{-1})^T {\bf F}\bar{y}'= {\bf \lambda}
\bar{q}'.
\end{equation}
Inserting Eq. (\ref{A2G}), gives the eigenvalue equation for the
normal mode coordinate $\bar{q}'$:
\begin{equation}
{\bf G} {\bf F}\bar{q}' = \lambda \bar{q}'.
\end{equation}
Thus, the FG matrix method consists of finding the roots of the
characteristic polynomial in $\lambda$:
\begin{equation}
\label{GF} \det(\lambda{\bf I}-{\bf G}{\bf F})=0,
\end{equation}
which is carried out for a general $N$-body system in Sec.
\ref{sec:firstorder} and applied to the systems of Secs.
\ref{sec:atom} - \ref{sec:hs}.

\renewcommand{\theequation}{B\arabic{equation}}
\setcounter{equation}{0}
\section{Spectral graph theory and the symmetry of the Q matrices}
\label{app:graph} In this appendix, we introduce the relevant
aspects of graph theory in order to derive several quantities used
throughout the main text and to show the connection of graphs to
the large-$D$ symmetric configuration of $N$ particles. A more
complete treatment of the mathematics of graph theory with an
emphasis on chemical physics applications may be found in Ref.
\cite{cvetkovic}.

A graph ${\mathcal G}=(V,E)$ consists of a finite set of vertices
$\{v_1,v_2,\ldots,v_n\}\epsilon V({\mathcal G})$ and a set of not
necessarily distinct unordered pairs $\{(v_i,v_j),\forall(1 \le i
< j \le n) \}\epsilon E({\mathcal G})$ connecting the vertices to
form the corresponding edges.  A graph is complete on n vertices,
denoted by $K_n({\mathcal G})$, if every distinct pair of vertices
is connected by an edge; $K_n$ is sometimes called a simplex of n
points. Notice that $K_n$ contains $n(n-1)/2$ edges. Our
large-dimension symmetric minimum configuration of $N$ atoms forms
a simplex on $N$ vertices in $D$-dimensional space.  For example,
the $N=4$ simplex is a hypertetrahedron. The line graph
$L({\mathcal G})$ of a graph ${\mathcal G}$ is the graph whose
vertices correspond to the edges of ${\mathcal G}$ with two
vertices being adjacent if and only if the corresponding edges in
${\mathcal G}$ have a vertex in common.

{\bf Derivation of Eq. (\ref{GFsym}):}  The ${\bf Q}$ matrices
display a high degree of symmetry in the large-dimension
configuration, but it is not trivial to write down the elements of
${\bf GF}$ in terms of its constituents ${\bf G}$ and ${\bf F}$ as
we have done in Eq. (\ref{GFsym}).  We now derive this result in a
two-step process. The first step is to write the $(N+M)$-square
matrices ${\bf G}$ and ${\bf F}$, where $M=N(N-1)/2$, as block
matrices of the following form:
\begin{equation}
{\bf Q}=\left(\begin{array}{cc} {\bf Q}_1 & {\bf Q}_2 \\ {\bf Q}_3
& {\bf Q}_4
\end{array}\right),
\end{equation}
where the blocks, each of which will be discussed in turn, have
the same indical structure as the blocks of ${\bf y}{\bf y}^{T}$
in Eq. (\ref{eq:yTy}).

From Eq. (\ref{eq:GFsyma}), the $(N \times N)$ ${\bf Q}_1$ block
is composed of two distinct elements: the diagonal elements
$Q_a=F_{ii}$ and the off-diagonal elements $Q_b=Q_{ij}$ ($i\ne
j$). ${\bf Q}_1$ can be written in terms of the more basic
matrices ${\bf I}_N$ (the $(N \times N)$ identity matrix) and
${\bf J}_N$ (the $(N \times N)$ matrix consisting of all ones):
\begin{equation}
{\bf Q}_1 = (Q_a - Q_b) {\bf I}_N + Q_b {\bf J}_N.
\end{equation}
The matrix ${\bf J}_N$ contains all ones, so for the diagonal
parts of ${\bf Q}_1$ the $Q_b$'s cancel and one is correctly left
with $Q_a$.  The off-diagonal parts of ${\bf Q}_1$ correctly
become $Q_b$ since the off-diagonal of ${\bf I}_N$ is zero.
Specifically, taking the ${\bf Q}$ matrix to be ${\bf F}$, we find
\begin{equation}
{\bf F}_1 = (F_a - F_b) {\bf I}_N + F_b {\bf J}_N.
\end{equation}
One can write down the ${\bf G}_1$ block in similar fashion noting
that the off-diagonal entries of ${\bf G}_1$ are zero:
\begin{equation}
{\bf G}_1 = G_a {\bf I}_N.
\end{equation}

The $(N \times M)$ ${\bf Q}_2$ block according to Eq.
(\ref{eq:GFsyma}) is also composed of two distinct elements: the
``incident'' elements $Q_e=Q_{i,ij}$ ($i\ne j$) and the
``non-incident'' elements $Q_f=Q_{i,jk}$ ($i\ne j \ne k$).  The
elements $F_e$ are termed ``incident'' because the repeated index
$i$ means that the vertex $i$ is incident with edge $(ij)$, while
for $Q_f$ the vertex designated $i$ is not incident with the edge
$(jk)$ since $i\ne j \ne k$.  To write ${\bf Q}_2$ in terms of
basic matrices, we need the $(N \times M)$ vertex-edge matrix
${\bf R}$, defined as $R_{i,jk}=\delta_{ij}+\delta_{ik}$, which
equals one when vertex $i$ is incident with edge $(jk)$ (i.e.,
when $i=j$ or $i=k$) and zero otherwise. Thus, the following
equation accurately describes the ${\bf Q}_2$ block:
\begin{equation}\label{eq:Q2}
{\bf Q}_2 = (Q_e - Q_f) {\bf R} + Q_f {\bf J}_{NM}.
\end{equation}
Specifically, letting the ${\bf Q}$ matrix be ${\bf F}$, we find
\begin{equation}\label{F2}
{\bf F}_2 = (F_e - F_f) {\bf R} + F_f {\bf J}_{NM}.
\end{equation}
And since there is no mixing of radial and angular derivatives in
the Schr\"odinger equation (\ref{eq:SE}), we have
\begin{equation}
{\bf G}_2 = {\bf G}_3={\bf 0}.
\end{equation}

By definition, the Hessian matrix ${\bf F}$ is symmetric; thus,
from Eq. (\ref{F2}) we find
\begin{equation}
{\bf F}_3 = (F_e - F_f) {\bf R}^T + F_f {\bf J}_{MN},
\end{equation}
where ${\bf J}_{MN}={\bf J}^T_{NM}$.  However, for a more general
${\bf Q}$-matrix (e.g., ${\bf GF}$ which, unlike ${\bf F}$, is not
symmetric), an analysis analogous to that leading to Eq.
(\ref{eq:Q2}) yields
\begin{equation}
{\bf Q}_3 = (Q_c - Q_d) {\bf R}^T + Q_d {\bf J}_{MN},
\end{equation}
where $Q_c=Q_{ij,i}$ ($i\ne j$) and $Q_d=Q_{jk,i}$ ($i\ne j \ne
k$).   The ${\bf Q}_3$ block can be interpreted in the same was as
${\bf Q}_2$ where the ``incident'' elements are now $Q_c$ and the
``non-incident'' elements are $Q_d$

For the fourth block of ${\bf Q}$, we need to introduce the
adjacency matrix of the line graph of the simplex, i.e., the
quantity ${\bf R}^T{\bf R}-2{\bf I}_{M}$.  A more intuitive way to
interpret this quantity is to think of it as an edge-edge matrix
that is unity when two edges are adjacent and zero otherwise. For
the purposes of constructing the ${\bf Q}_4$ matrix, which has
indices of the form $(ij,kl)$, it helps to then think of the index
$(ij)$ as an edge of the simplex connecting vertices $i$ and $j$,
and likewise for edge $(kl)$. Two edges are adjacent if they share
a vertex. ${\bf Q}_4$ is comprised of three distinct elements: the
diagonal elements $Q_g=Q_{ij,ij}$ ($i \ne j$), the adjacent-edge
elements $Q_h=Q_{ij,jk}$ ($i \ne j \ne k$), and the
non-adjacent-edge elements $Q_{\iota}=Q_{ij,kl}$ ($i \ne j \ne k
\ne l$), which have no repeated indices and, therefore, no
adjacent edges. The matrix ${\bf R}^T{\bf R}$, whose
$(ij,kl)^{th}$ element is given by
$I_{ik}(J_N)_{jl}+I_{il}(J_N)_{jk}+I_{jk}(J_N)_{il}+I_{jl}(J_N)_{ik}$,
gives the adjacent edges, but it double counts the diagonal; thus,
the term $Q_h({\bf R}^T{\bf R}-2{\bf I}_{M})$ of ${\bf Q}_4$
accounts for the adjacent edges. To get the diagonal elements of
${\bf Q}_4$, we need the term $Q_g {\bf I}_M$.  And for the
non-adjacent edges we need $Q_{\iota}[{\bf J}_M-({\bf R}^T{\bf
R}-{\bf I}_M)]$. Putting these terms together gives:
\begin{equation}
{\bf Q}_4 = (Q_g - 2Q_h+ Q_{\iota}){\bf I}_M+(Q_h-Q_{\iota}) {\bf
R}^T{\bf R} + Q_{\iota} {\bf J}_{M}.
\end{equation}
Specifically, one has
\begin{equation}
{\bf F}_4 = (F_g - 2F_h+ F_{\iota}){\bf I}_M+(F_h-F_{\iota}) {\bf
R}^T{\bf R} + F_{\iota} {\bf J}_{M},
\end{equation}
and likewise for ${\bf G}_4$, noting that $G_{\iota}=0$:
\begin{equation}
{\bf G}_4 = (G_g - 2G_h){\bf I}_M+ G_h {\bf R}^T{\bf R}.
\end{equation}

The second step to derive Eq. (\ref{GFsym}) is to find the blocks
of ${\bf GF}$ by matrix multiplication of ${\bf G}$ and ${\bf F}$,
whose blocks are given above. That is, we multiply the following
\begin{equation}
{\bf G}=\left(\begin{array}{cc} {\bf G}_1 & {\bf 0} \\ {\bf 0} &
{\bf G}_4
\end{array}\right)\;\;\;\;\; {\bf F}=\left(\begin{array}{cc}
{\bf F}_1 & {\bf F}_2 \\
{\bf F}_3 & {\bf F}_4
\end{array}\right).
\end{equation}
Then by analyzing each block of ${\bf GF}$ we can write down its
elements in terms of the elements of ${\bf G}$ and ${\bf F}$. The
first block of ${\bf GF}$ gives $a$ and $b$ of Eq. (\ref{GFsym})
and is given by
\begin{equation}
({\bf GF})_1 = {\bf G}_1 {\bf F}_1 = (G_a F_a -  G_a F_b){\bf I}_N
+ G_a F_b {\bf J}_N.
\end{equation}
Taking the diagonal part of $({\bf GF})_1$ gives
\begin{equation}
a \equiv (GF)_{a} \equiv (GF)_{i,i} = G_a F_a,
\end{equation}
while the off-diagonal $(i \ne j)$ gives
\begin{equation}
b \equiv (GF)_{b} \equiv (GF)_{i,j} = G_a F_b.
\end{equation}

The second block gives $e$ and $f$ of Eq. (\ref{GFsym}) and is
given by
\begin{equation}
({\bf GF})_2 = {\bf G}_1 {\bf F}_2 = (G_a F_e -  G_a F_f){\bf R} +
G_a F_f {\bf J}_{NM}.
\end{equation}
Noting that $R_{i,ij}$ is unity when $i \ne j$ and that all
elements of ${\bf J}_{NM}$ are unity, we find the following from
$({\bf GF})_2$ when $i \ne j$:
\begin{equation}
e \equiv (GF)_{e} \equiv (GF)_{i,ij} = G_a F_e.
\end{equation}
For $i \ne j \ne k$, $R_{i,jk}=0$; and so,
\begin{equation}
f \equiv (GF)_{f} \equiv (GF)_{i,jk} = G_a F_f.
\end{equation}

When deriving $({\bf GF})_3$ and $({\bf GF})_4$ below, the
following relations prove useful:
\begin{eqnarray}
{\bf R}{\bf R}^T = {\bf J}_N+(N-2){\bf I}_N & {\bf R}^T {\bf
J}_{N}=2 {\bf J}_{MN} & {\bf R} {\bf J}_{MN}=(N-1) {\bf J}_{N} \\
{\bf R} {\bf J}_{M}=(N-1) {\bf J}_{NM} & {\bf R}^T {\bf J}_{NM}= 2
{\bf J}_{M} & .
\end{eqnarray}
Equations $c$ and $d$ of Eq. (\ref{GFsym})
can be found from the third block of ${\bf GF}$, which is given by
\begin{eqnarray}
({\bf GF})_3 = {\bf G}_4 {\bf F}_3 = &\left[G_g F_e - G_g F_f +
(N-4) (G_h F_e - G_h F_f)\right]{\bf R}^T& \nonumber \\
\label{gf3} & + \left[ G_g F_f + 2 G_h F_e + 2(N-3) G_h F_f
\right] {\bf J}_{MN}.&
\end{eqnarray}
For the $(ij,i)$ elements $(i \ne j)$, ${\bf R}^T$ is unity; thus,
to get $(GF)_{c}$ one adds both bracketed terms in Eq.
(\ref{gf3}):
\begin{eqnarray}
c \equiv (GF)_{c} \equiv (GF)_{ij,i} = G_g F_e + (N-2) G_h F_e +
(N-2) G_h F_f.
\end{eqnarray}
For the $(jk,i)$ elements $(i \ne j \ne k)$, ${\bf R}^T$ is zero;
thus, $(GF)_{d}$ is simply given by the second bracketed term in
Eq. (\ref{gf3}):
\begin{eqnarray}
d \equiv (GF)_{d} \equiv (GF)_{jk,i} = G_g F_f + 2 G_h F_e +
2(N-3) G_h F_f.
\end{eqnarray}

Equations $g$, $h$, and $\iota$ of Eq. (\ref{GFsym}) can be found
from the fourth block of ${\bf GF}$, which is given by
\begin{eqnarray} ({\bf GF})_4 = {\bf G}_4 {\bf F}_4 = &\left[(G_g -
2G_h)(F_g - 2 F_h + F_i)\right]{\bf I}_M + \left[
G_g F_{\iota} + 4G_h F_h + 2(N-4) G_h F_{\iota} \right] {\bf J}_{M}& \nonumber \\
& \label{gf4} + \left[ (N-6) G_h F_h - (N-5) G_h F_{\iota} + G_g
(F_h - F_{\iota}) + G_h F_g \right] {\bf R}^T {\bf R}. &
\end{eqnarray}
As noted earlier, ${\bf R}^T {\bf R}$ double counts the diagonal
elements (i.e., $(ij,ij)$ elements where $i \ne j$). Thus, the
$(GF)_{ij,ij}$ elements are given by the sum of the first two
bracketed terms in Eq. (\ref{gf4}) plus twice the third term:
\begin{eqnarray}
g \equiv (GF)_{ij,ij} \equiv (GF)_{ij,ij} = G_g F_g + 2 (N-2) G_h
F_h.
\end{eqnarray}
For the $(ij,jk)$ elements $(i \ne j \ne k)$, we note that the
${\bf R}^T {\bf R}$ elements are unity and the elements of ${\bf
I}_N$ are zero. Thus, to get $(GF)_{ij,jk}$ we take the sum of the
second and third bracketed term in Eq. (\ref{gf4}):
\begin{eqnarray} h \equiv (GF)_{h} \equiv (GF)_{ij,jk}
= G_g F_h +  G_h F_g + (N-2) G_h F_h + (N-3) G_h F_{\iota}.
\end{eqnarray}
Finally, the $(ij,kl)$ elements $(i \ne j \ne k \ne l)$ of ${\bf
R}^T {\bf R}$ and ${\bf I}_N$ are zero, and thus the last equation
in Eq. (\ref{GFsym}) is simply given by the second bracketed term in
Eq. (\ref{gf4}):
\begin{eqnarray} \iota \equiv (GF)_{\iota} \equiv (GF)_{ij,kl}
= G_g F_{\iota} + 4G_h F_h + 2(N-4) G_h F_{\iota}.
\end{eqnarray}

{\bf Derivation of Eq. (\ref{eq:detalg}):} A graph with a given
structure corresponds to a spectrum. That is, a graph may be
represented by a matrix whereby its spectrum, or eigenvalues, may
be calculated.  $P_{\mathcal G}(\lambda)$ denotes the
characteristic polynomial of the adjacency matrix ${\bf A}$ of
graph ${\mathcal G}$:
\begin{equation}
P_{\mathcal G}(\lambda)\equiv \det(\lambda {\bf I}-{\bf A}).
\end{equation}
The element $(i,j)$ of the adjacency matrix is the number of edges
connecting vertices i and j of the graph ${\mathcal G}$.  The
spectrum of ${\mathcal G}$ is found by solving $P_{\mathcal
G}(\lambda)=0$.

To prove the relationships involving determinants in this appendix
and the next, we quote the following result from p.72 of Ref.
\cite{cvetkovic} for the complete graph $K_n$:
\begin{equation}\label{PKn}
P_{K_n}(\lambda)=(\lambda-n+1)(\lambda+1)^{n-1}.
\end{equation}
The adjacency matrix of a simplex of n points, $K_n$, is an $n
\times n$ matrix of ones except along the diagonal which contains
all zeros. That is, $P_{K_n}(\lambda)$ takes the following form:
\begin{equation}\label{eq:pkn2}
P_{K_n}(\lambda) = \det \left(
\begin{array}{cccc}
\lambda & -1        & \cdots & -1      \\
-1        & \lambda &  \ddots        & \vdots   \\
\vdots  &  \ddots          & \ddots &  -1      \\
-1        & \cdots  &   -1     & \lambda \\
\end{array}
\right).
\end{equation}
Using Eq. (\ref{PKn}) one can easily derive Eq. (\ref{eq:detalg})
of Sec. \ref{sec:firstorder} , which states
\begin{equation}
\det({\bf XYEZ })=\det \left(\begin{array}{cc} \frac{t}{v} {\bf
I}_N + \frac{u}{v} {\bf J}_N & {\bf 0} \\ \cdots & v {\bf I}_M
\end{array}\right)= t^{N-1} (t + N u) v^{M-N},
\end{equation}
where the integer $M$ is defined in Eq. (\ref{eq:M}).  First we
note that
\begin{equation}
\det \left(\begin{array}{cc} \frac{t}{v} {\bf I}_N + \frac{u}{v}
{\bf J}_N & {\bf 0} \\ \cdots & v {\bf I}_M
\end{array}\right)= \det\left(v {\bf I}_M \right) \det\left(\frac{t}{v} {\bf I}_N + \frac{u}{v}
{\bf J}_N \right).
\end{equation}
Then, $\det(\frac{t}{v} {\bf I}_N + \frac{u}{v} {\bf J}_N)$ can be
written as an $(N \times N)$ matrix of the same form as
$P_{K_n}(\lambda)$ in Eq. (\ref{eq:pkn2}):
\begin{equation}\label{PKnMatrix}
\det(\frac{t}{v} {\bf I}_N + \frac{u}{v} {\bf J}_N) = \det \left(
\begin{array}{cccc}
\frac{t+u}{v} & \frac{u}{v}   & \cdots & \frac{u}{v}      \\
\frac{u}{v}        & \frac{t+u}{v} &  \ddots        & \vdots   \\
\vdots  &  \ddots          & \ddots &  \frac{u}{v}      \\
\frac{u}{v}        & \cdots  &   \frac{u}{v}     & \frac{t+u}{v} \\
\end{array}
\right) = \left( -\frac{u}{v}\right)^N \det \left(
\begin{array}{cccc}
-1-\frac{t}{u} & -1        & \cdots & -1      \\
-1        & -1-\frac{t}{u} &  \ddots        & \vdots   \\
\vdots  &  \ddots          & \ddots &  -1      \\
-1        & \cdots  &   -1     & -1-\frac{t}{u} \\
\end{array}
\right),
\end{equation}
so that, noting $\det(v {\bf I}_M)=v^M$, and setting
$\lambda=-1-\frac{t}{u}$ and $n=N$ in Eq. (\ref{PKn}) leads
directly to $\det({\bf XYEZ })=t^{N-1}(t + N u) v^{M-N}$.

\renewcommand{\theequation}{C\arabic{equation}}
\setcounter{equation}{0}
\section{Gaussian Reduction of Equation (\ref{eq:ematrix})}\label{app:xyez}
The characteristic equation for the ${\bf GF}$ eigenvalues is
$\det({\bf E})=0$, where ${\bf E}$ is given by Eq.
(\ref{eq:ematrix}).  The Gaussian elimination to reduce ${\bf E}$
of Eq. (\ref{eq:ematrix}) to the form ${\bf XYEZ}$ of Eq.
(\ref{xyez}) allows us to find analytical expressions for the
normal-mode frequencies.   The goal is to transform ${\bf E}$ to a
lower-triangular matrix, whose determinant we can compute exactly
with Eq. (\ref{eq:detalg}), while leaving the characteristic
determinant invariant by imposing that the transforming matrices
have unit determinant. The elimination process consists of three
steps.

{\bf Step 1:} We first define ${\bf Z}$:
\begin{equation}
{\bf Z}\equiv  \left(\begin{array}{cc} {\bf I}_N & w {\bf J}_{NM}
\\ {\bf 0} & {\bf I}_M
\end{array}\right),
\end{equation}
whose determinant is unity, which we now show.  It is known from
matrix theory (for example, see Ref. \cite{marcus}) that the
determinant of an ($N+M$)-square matrix with an ($N \times M$)
zero matrix (${\bf 0}_{N,M}$) in the upper right block or an ($M
\times N$) zero matrix (${\bf 0}_{M,N}$) in the lower left block
is the product of the determinant of the diagonal block of
matrices. That is,
\begin{equation}\label{eq:det1}
\det  \left(\begin{array}{cl} {\bf A} & {\bf 0}_{NM}
\\ {\bf C} & {\bf B}
\end{array}\right)=\det  \left(\begin{array}{lc} {\bf A} & {\bf C}
\\ {\bf 0}_{MN} & {\bf B}
\end{array}\right)=\det({\bf A})\det({\bf B}),
\end{equation}
where ${\bf A}$ and ${\bf B}$ are square matrices. For ${\bf Z}$,
we have $\det({\bf A})=\det({\bf I}_N)=1$ and $\det({\bf
B})=\det({\bf I}_M)=1$; thus, $\det({\bf Z})=1$.

Multiplying ${\bf E}$ on the right by ${\bf Z}$ and using the
multiplication rules above leads to
\begin{equation}
{\bf EZ} = \left(\begin{array}{cc}
(\lambda-a+b){\bf I}_N - b {\bf J}_N & (f-e) {\bf R}+[w(\lambda-a-(N-1)b)-f] {\bf J}_{NM} \\
(d-c){\bf R}^T - d{\bf J}^T_{NM} & (\lambda-g+2h-\iota){\bf I}_M -
(h-\iota){\bf R}^T{\bf R} + [(2(d-c)- Nd)w -\iota]{\bf J}_M
\end{array}\right).
\end{equation}
We choose $w$ equal to $f/(\lambda-a-(N-1)b)$, such that the
upper-right ${\bf J}_{NM}$ term vanishes. This results in:
\begin{equation}
{\bf EZ} = \left(\begin{array}{cc}
(\lambda-a+b){\bf I}_N - b {\bf J}_N & (f-e) {\bf R} \\
(d-c){\bf R}^T - d{\bf J}^T_{NM} & v{\bf I}_M - (h-\iota){\bf
R}^T{\bf R} + \zeta{\bf J}_M
\end{array}\right),
\end{equation}
where
\begin{eqnarray}
\zeta \equiv f\left(\frac{2(d-c)-Nd}{\lambda-a-(N-1)b} \right)-\iota \\
v \equiv \lambda-g+2h-\iota.
\end{eqnarray}

{\bf Step 2:} Next multiply on the left by ${\bf Y}$, which by Eq.
(\ref{eq:det1}) has unit determinant:
\begin{equation}
{\bf Y} \equiv \left(\begin{array}{cc} {\bf I}_N & {\bf 0}\\
x {\bf R}^T + y {{\bf J}_{NM}}^T & {\bf I}_M
\end{array}\right),
\end{equation}
where we impose
\begin{equation}
x=\frac{h-i}{f-e}  \;\;\;\;\; y=-\frac{\zeta}{2
(f-e)}=-\frac{1}{2(f-e)}\left[f \left(
\frac{2(d-c)-Nd}{\lambda-a-(N-1)b}\right)-\iota\right]
\end{equation}
to eliminate the lower-right ${\bf R}^T{\bf R}$ and ${\bf J}_M$
terms from ${\bf EM}$.  This results in
\begin{equation}
{\bf YEZ} = \left(\begin{array}{cc}
(\lambda-a+b){\bf I}_N - b {\bf J}_N & (f-e) {\bf R} \\
l{\bf R}^T + m {\bf J}^T_{NM} & v {\bf I}_M
\end{array}\right),
\end{equation}
where
\begin{eqnarray}
l &\equiv&\frac{h-\iota}{f-e}(\lambda-a+b)+(d-c) \nonumber \\
m &\equiv& y(\lambda-a-(N-1)b)-2b\frac{h-\iota}{f-e}-d \\
\nonumber
  &=& -\frac{1}{2(f-e)}[f(2(d-c)-Nd)-\iota(\lambda-a-(N-1)b)]-2b\frac{h-\iota}{f-e}-d.
\end{eqnarray}

{\bf Step 3:} The final step is to eliminate the upper right term
from ${\bf YEZ}$ by multiplying ${\bf YEZ}$ by ${\bf X}$, which by
Eq. (\ref{eq:det1}) has unit determinant:
\begin{equation}
{\bf X} \equiv \left(\begin{array}{cc} {\bf I}_N & z{\bf R}\\
{\bf 0} & {\bf I}_M
\end{array}\right),
\end{equation}
where $z=(e-f)/v$.  {\bf XYEZ} results in the desired form of Eq.
(\ref{xyez}):
\begin{equation}
{\bf XYEZ} = \left(\begin{array}{cc}
\frac{t}{v} {\bf I}_N + \frac{u}{v} {\bf J}_N & {\bf 0} \\
l{\bf R}^T + m {\bf J}^T_{NM} & v {\bf I}_M
\end{array}\right),
\end{equation}
where
\begin{eqnarray}
t&=&(\lambda-a+b)v + (N-2)(\iota-h)(\lambda-a+b) + (N-2)(d-c)(e-f)
\\ u &=&
-kv-(h-\iota)(\lambda-a+b)+(d-c)(e-f) \nonumber \\
&\; \;& +\frac{N-1}{2}
\left[f(2(d-c)-Nd)-\iota(\lambda-a-(N-1)b)+4b(h-\iota)-2d(e-f)\right].
\end{eqnarray}

As the determinants of ${\bf Z}$, ${\bf Y}$, and ${\bf X}$ are
unity, the determinant of ${\bf XYEZ}$ and, hence, the
characteristic determinant $\det({\bf E})$ may be calculated using
Eq. (\ref{eq:detalg}), which was derived in the previous appendix.

\renewcommand{\theequation}{D\arabic{equation}}
\setcounter{equation}{0}
\section{Gramian Determinants}\label{app:gram}
The Gramian determinant\cite{gantmacher} is defined as:
\begin{equation}
\Gamma\equiv\det(\gamma_{ij}),
\end{equation}
where $\gamma_{ij}={\bf r}_{i}\cdot{\bf r}_{j}/r_i r_j$, the angle
cosines between the particle radii ${\bf r}_{i}$ and ${\bf
r}_{j}$, represents the elements of an $N \times N$ matrix.  A
related quantity used in the main text is the principle minor of
the Gramian, $\Gamma^{(\alpha)}$, defined as the determinant of
the $\gamma_{ij}$ matrix with the $\alpha^{th}$ row and column
removed.

A most challenging part of calculating the large-$D$ minimum and
the ${\bf F}$ matrix elements in the systems discussed in this
paper is handling the Gramian determinants and their derivatives.
What makes these calculations feasible is the very high symmetry
of the infinite-dimension, symmetric minimum.  We make use of Eq.
(\ref{PKn}) to obtain the Gramian determinant and its derivatives
at the infinite-dimension, symmetric minimum. We will now
demonstrate how this is done and summarize the results.

In general $\gamma_{ii}=1$ and at the infinite-$D$ symmetric
minimum all of the remaining direction cosines are equal,
$\gamma_{ij}=\gamma_{\infty}$.  Hence, the Gramian determinant is
an $(N \times N)$ matrix of the form
\begin{equation}
\Gamma \Big|_{\infty} = \det \left(
\begin{array}{cccc}
1 & \gamma_{\infty}        & \cdots & \gamma_{\infty}      \\
\gamma_{\infty}        & 1 &  \ddots        & \vdots   \\
\vdots  &  \ddots          & \ddots &  \gamma_{\infty}      \\
\gamma_{\infty}        & \cdots  &   \gamma_{\infty}     & 1 \\
\end{array}
\right),
\end{equation}
which can be written in the same form as $P_{K_n}(\lambda)$ (Eq.
(\ref{eq:pkn2}) of Appendix B):
\begin{equation}
\Gamma\Big|_{\infty} = \left(-\gamma_{\infty} \right)^N \det
\left(
\begin{array}{cccc}
-1/\gamma_{\infty} & -1        & \cdots & -1      \\
-1        & -1/\gamma_{\infty} &  \ddots        & \vdots   \\
\vdots  &  \ddots          & \ddots &  -1      \\
-1        & \cdots  &   -1     & -1/\gamma_{\infty} \\
\end{array}
\right).
\end{equation}
Setting $\lambda=-1/\gamma_{\infty}$ and $n=N$ in Eq. (\ref{PKn})
gives
\begin{equation}\label{eq:gammainf}
\Gamma\Big|_{\infty} =
[1+(N-1)\gamma_{\infty}](1-\gamma_{\infty})^{N-1}.
\end{equation}
The principle minor evaluated at the infinite-$D$ symmetric
minimum $\Gamma^{(\alpha)}\Big|_{\infty}$ is simply related to the
corresponding Gramian determinant (Eq. (\ref{eq:gammainf})) by $N \to
N-1$.

To calculate the infinite-$D$ symmetric minimum of the Gramian
derivatives, we expand $\Gamma$ in terms of its cofactors.  The
cofactor, denoted by $C_{ij}$, of the element $\gamma_{ij}$ in
$\Gamma$ is $(-1)^{i+j}$ multiplied by the determinant of the
matrix obtained by deleting the $i^{th}$ row and $j^{th}$ column
of $\Gamma$.  We may then write $\Gamma$ as
\begin{equation}
\Gamma=\sum_{j=1}^N \gamma_{ij} C_{ij}.
\end{equation}
Then the partial derivative of $\Gamma$ in terms of the cofactor
is
\begin{equation}
\frac{\partial \Gamma}{\partial \gamma_{ij}}=2C_{ij}.
\end{equation}
From this equation, the partial derivative of $\Gamma$ evaluated
at the infinite-$D$ symmetric minimum is
\begin{equation}
\left. \frac{\partial \Gamma}{\partial
\gamma_{ij}}\right|_{\infty}=-2 C_{\infty}^{(N-1)},
\end{equation}
where we have defined the following determinant of an $(N-1)
\times (N-1)$ matrix:
\begin{equation}
C_{\infty}^{(N-1)}=\det \left(
\begin{array}{ccccc}
\gamma_{\infty}& \gamma_{\infty}&\gamma_{\infty}& \cdots & \gamma_{\infty}\\
\gamma_{\infty} & 1 &  \gamma_{\infty}&  & \vdots \\
\gamma_{\infty}&  \gamma_{\infty}& 1 & \ddots & \vdots    \\
\vdots & \ddots & \ddots & \ddots & \gamma_{\infty} \\
\gamma_{\infty} & \cdots    & \cdots   &   \gamma_{\infty}     & 1 \\
\end{array}
\right),
\end{equation}
and the superscript $(N-1)$ simply indicates the size of the
matrix. From this matrix one can show that the following recursion
relation holds:
\begin{equation}
C_{\infty}^{(N-1)}=\gamma_{\infty}\left(
\Gamma\Big|^{(N-2)}_{\infty} - (N-2) C_{\infty}^{(N-2)} \right),
\end{equation}
or equivalently
\begin{equation}\label{eq:recursion}
C_{\infty}^{(N)}=\gamma_{\infty}\left(
\Gamma\Big|^{(N-1)}_{\infty} - (N-1) C_{\infty}^{(N-1)} \right),
\end{equation}
where the $(N)$ superscript in the notation
$\Gamma\Big|^{(N)}_{\infty}$ again refers to the size of the
matrix $\Gamma\Big|_{\infty}$. From the recursion relation
(Eq. (\ref{eq:recursion})), one can easily prove by induction the
conjecture that
$C_{\infty}^{(N)}=\gamma_{\infty}(1-\gamma_{\infty})^{N-2}$ and
so,
\begin{equation}\label{eq:gamderiv}
\left. \frac{\partial \Gamma}{\partial
\gamma_{ij}}\right|_{\infty}=-2
\gamma_{\infty}(1-\gamma_{\infty})^{N-2}.
\end{equation}
The derivative of the principle minor evaluated at the
infinite-$D$ symmetric minimum is simply related to the
corresponding Gramian determinant derivative (Eq. (\ref{eq:gamderiv}))
by $N \to N-1$.

To summarize the above results, the following expressions are
needed when calculating the minimum of the effective potential
(Eqs. (\ref{minimum1}) and (\ref{minimum2})):
\begin{equation}\label{firstgammas}
\begin{array}{ll}
\left. \frac{\partial{\Gamma}}{\partial
\gamma_{ij}}\right|_{\infty}=-2\gamma_{\infty}(1-\gamma_{\infty})^{N-2}
& \left. \frac{\partial{\Gamma^{(\alpha)}}}{\partial
\gamma_{ij}}\right|_{\infty}=-2\gamma_{\infty}(1-\gamma_{\infty})^{N-3}
\\
\left.
\Gamma\right|_{\infty}=[1+(N-1)\gamma_{\infty}](1-\gamma_{\infty})^{N-1}
& \left.
\Gamma^{(\alpha)}\right|_{\infty}=[1+(N-2)\gamma_{\infty}](1-\gamma_{\infty})^{N-2}.
\end{array}
\end{equation}

And when evaluating the ${\bf F}$ matrix elements at the
infinite-$D$ symmetric minimum, the following six second-order
derivatives of the Gramian determinants are needed:
\begin{equation}\label{secondgammas}
\begin{array}{lll}
\left. \frac{\partial^2{\Gamma}}{\partial \gamma_{ij}\partial
\gamma_{kl}}\right|_{\infty}=0 & \left.
\frac{\partial^2{\Gamma}}{\partial \gamma_{ij}^2}\right|_{\infty}
= -2(1-\gamma_{\infty})^{N-3}(1+(N-3)\gamma_{\infty}) & \left.
\frac{\partial^2{\Gamma}}{\partial \gamma_{ij}\partial
\gamma_{jk}}\right|_{\infty}=2\gamma_{\infty}(1-\gamma_{\infty})^{N-3} \\ \\
\left. \frac{\partial^2{\Gamma^{(\alpha)}}}{\partial
\gamma_{ij}\partial \gamma_{kl}}\right|_{\infty}=0 & \left.
\frac{\partial^2{\Gamma^{(\alpha)}}}{\partial
\gamma_{ij}^2}\right|_{\infty} =
-2(1-\gamma_{\infty})^{N-4}(1+(N-4)\gamma_{\infty}) & \left.
\frac{\partial^2{\Gamma^{(\alpha)}}}{\partial \gamma_{ij}\partial
\gamma_{jk}}\right|_{\infty} =
2\gamma_{\infty}(1-\gamma_{\infty})^{N-4}.
\end{array}
\end{equation}


\begin{thebibliography}{99}
\bibitem{loeser} J.\ G.\ Loeser, J.\ Chem.\ Phys.\ \textbf{86}, 5635 (1987).
\bibitem{dcw} E.\ B.\ Wilson, Jr., J.\ C.\ Decius, P.\ C.\ Cross,
\textit{Molecular vibrations: The theory of infrared and raman
vibrational spectra}. McGraw- Hill, New York, 1955.
\bibitem{avery}J.\ Avery, D.\ Z.\ Goodson, D.\ R.\ Herschbach,
Theor.\ Chim.\ Acta \textbf{81}, 1 (1991).
\bibitem{matrix_method} M.\ Dunn, T.\ C.\ Germann, D.\ Z.\ Goodson, C.\
A.\ Traynor, J.\ D.\ Morgan III, D.\ K.\ Watson, and D.\ R.\ Herschbach,
J.\ Chem.\ Phys.\ {\bf 101} 5987 (1994).
\bibitem{jacobian} M.\ Dunn, B.\ A.\ McKinney, and D.\ K.\ Watson, (to be submitted).
\bibitem{chat} A.\ Chatterjee, J.\ Phys.\ A: Math.\ Gen.\ {\bf 18}, 735
(1985).
\bibitem{strang} G.\ Strang, {\it Linear algebra and its
applications, Third ed.}. Harcourt Brace Jovanovich College
Publishers, Orlando, FL, 1988.
\bibitem{different} In Ref.~\cite{loeser}, Eq. (\ref{eq:E1}) would read
%
\begin{eqnarray}
\overline{E} &=& \overline{E}_{\infty} + \delta \overline{E}_o + O(\delta^2) \nonumber \\
&=&V_{\mathtt{eff}}(\bar{r}_{\infty},\gamma_{\infty}) +
\hspace{0.50em} \delta
\hspace{-2.00em}
\sum_{\renewcommand{\arraystretch}{0}
\begin{array}[t]{r@{}l@{}c@{}l@{}l} \scriptstyle \mu = \{
  & \scriptstyle \mathbf{0}^\pm,\hspace{0.5ex}
  & \scriptstyle \mathbf{1}^\pm & , & \\
  & & \scriptstyle \mathbf{2} & & \scriptstyle  \}
            \end{array}
            \renewcommand{\arraystretch}{1} }
\hspace{-1.5em}
( n_{\mu}+\frac{d_\mu}{2} ) \, \bar{\omega}_{\mu} +
O(\delta^2) \,,
\end{eqnarray}
%
where $n_{\mu}$ is the total number of quanta in all the normal
modes with the same frequency $\bar{\omega}_{\mu}$, i.e.\
%
\begin{equation}
n_\mu = \sum_{\mathsf{n}_{\mu}=0}^\infty
           {\mathsf{n}}_{\mu} \, d_{\mu,\mathsf{n}_{\mu}} \,.
\end{equation}
%
\bibitem{hamermesh} M.\ Hamermesh, {\it Group theory and its
application to physical problems}. Addison-Wesley, Reading, MA,
1962
\bibitem{cvetkovic}  D.\ M.\ Cvetkovi\'c, M.\ Doob, H.\ Sachs,
\textit{Spectra of graphs: Theory and application}. Academic
Press, New York, 1980.
\bibitem{marcus} M.\ Marcus and H.\ Ming, {\it A survey of matrix theory and matrix
inequalities}. Allyn and Bacon, Inc., Boston, 1964.
\bibitem{ashoori} R.\ C.\ Ashoori, Nature {\bf 379}, 413 (1996).
\bibitem{qdot} B.\ A.\ McKinney and D.\ K.\ Watson, Physical Review B
{\bf 61}, 4958 (2000).
\bibitem{baym} G.\ Baym and C.\ J.\ Pethick, Physical Review Letters,
{\bf 76}, 6 (1996).
\bibitem{blume} D.\ Blume and C.\ H.\ Greene, Physical Review A {\bf
63}, 63061 (2001).
\bibitem{mbbecpaper} B.\ A.\ McKinney, M.\ Dunn, and D.\ K.\ Watson, (to be submitted).
\bibitem{becreview} F.\ Dalfovo, S.\ Giorgini, L.\ P.\ Pitaevskii, and
S.\ Stringari, Reviews of Modern Physics {\bf 71}, 463 (1999).  A.\
J.\ Leggett, Reviews of Modern Physics {\bf 73}, 307 (2001).
\bibitem{dptgp} B.\ A.\ McKinney and D.\ K.\ Watson, Physical Review A
{\bf 65}, 33604 (2002).
\bibitem{gantmacher} F.\ R.\ Gantmacher, {\it The theory of
matrices, Vol.\ 1}. Chelsea, New York, 1959.
\bibitem{loeser2} Z.\ Zhen and J.\ G.\ Loeser, {\it Dimensional scaling
in chemical physics}. Ed.\ D.\ R.\ Herschbach, J.\ Avery, and O.\
Goscinski. Kluwer, Dordrecht, 1992. 83-114.
\end{thebibliography}
\end{document}